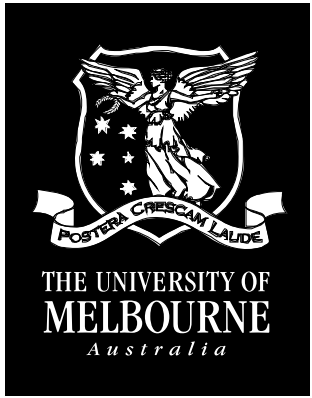

# Construction and Analysis of a Simplified Many-Body Neutrino Model

Ivona Okuniewicz

Submitted in total fulfillment of the requirements

of the degree of

Doctor of Philosophy

School of Physics

The University of Melbourne

Australia

January 2006

# Abstract

## Construction and Analysis of a Many-Body Neutrino model

by Ivona Okuniewicz

Principal Supervisor: Bruce H.J. Mckellar

In systems such as the early universe and supernovae neutrinos comprise a large fraction of the total particle number density thus one needs to consider neutrino self-refraction. Coherent neutrino-neutrino scattering has been found to play a role in the flavour evolution of the system. Traditionally this problem has been analysed by assuming that the wavefunction of the system can be factorised into one-body states. However in 1992 Pantaleone [1, 2] showed that a neutrino ensemble is in general a many-body problem due to the off-diagonal contribution to the neutrino refractive index. This topic was unexamined until recently. It has been suggested by Bell et. al. [3] that quantum entanglement could play an essential role in the flavour evolution of the dense neutrino system.

In this thesis we examine the validity of the one-body approximation by constructing a many-body neutrino model. The neutrino system is modelled by a system of interacting spins following earlier work [4]. We extend this work by generalising the model to initial states with asymmetric flavour composition. We find an exact analytical solution to the system. The investigation has revealed an array of interesting physics including semi-classical behaviour, quantum equilibration and a transition from semi-classical to purely quantum regimes. Further, this study has found no evidence for the violation of the one-body description of a dense neutrino ensemble.

We also note that our analysis is valid for any two state system with equal strength interactions.

# Declaration

This is to certify that

(i) the thesis comprises only my original work towards the PhD except where indicated in the Preface,

(ii) due acknowledgement has been made in the text to all other material used,

(iii) the thesis is less than 100,000 words in length, exclusive of tables, maps, bibliographies and appendices

Date .................... Signature ...................................................

# Preface

In the course of the PhD I have collaborated with Alexander Friedland and as a consequence two of the equations in this thesis, Eq.(4.13) and Eq.(4.63), were derived by Alexander Friedland. These equations are acknowledged as being found by Alexander Friedland in the thesis.

Further, a number of oral presentations have been made on the work in this thesis. Any conference proceedings are listed below.

## Conference Proceedings

I. Okuniewicz, B. McKellar, A. Friedland, *Construction and Analysis of a Many-Body Neutrino Model*, AIP 16th Biennial Congress, Canberra, Australia, Jan 31-Feb 4, 2005.

# Acknowledgements

I would like to thank Prof. Bruce H.J. McKellar for his guidance and patience during the term of the PhD research project. I would also like to acknowledge the helpful discussions with Alexander Friedland and Nicole F. Bell which have aided in my understanding of previous and current work on the dense neutrino system.

Finally, I would like to thank Gene for his emotional support and for working hard while I lived the Utopian life of a university student.

# Contents

# List of Figures

# Chapter 1

# Introduction

## 1.1 The conception of the neutrino

The early 20th century was dominated by nuclear physics and hence it is not a surprise that the neutrino was born due to a problem in nuclear physics. To understand the journey that led to the discovery of the neutrino we begin in the late 1800's. In December 1895, Roentgen observed X-rays and therefore propelled research into radioactivity. Four months later (March 1896) radioactivity was discovered by Becquerel. The short time between the two events was not an accident. Having heard about the discovery of X-rays, Becquerel developed an interest in the origin of X-rays. He wondered whether phosphorescent bodies emitted X-rays or similar rays. He then performed an experiment to search for these rays using uranium salts which was subsequently published in a paper titled "On radiations emitted in phosphorescence". In 1898 Marie Curie entered the arena with a paper that showed radioactivity is a property of individual atoms. This realisation led to the discovery of radium by Marie and Pierre Curie also in 1898. Then in 1899 Rutherford published a paper which showed that there were two distinct types of radiation which he called $\alpha$-radiation and $\beta$-radiation. It is the $\beta$-radiation which has played a significant role in the history of the neutrino.

The road that led to an understanding of $\beta$-decay was a long and winding one. The discovery of $\beta$-radiation and the early experiments on the $\beta$ spectrum preceded the postulation or observation of the neutron, proton and neutrino, and an understanding of the structure of an atom. Hence it is understandable that at this time the prediction for the $\beta$ spectrum was only based upon the energy spectrum of $\alpha$-decay which was found experimentally in 1904 to be monochromatic. The first few experiments on the $\beta$ spectrum seemed to confirm this prediction but this was due to a limited understanding (and in the case of the very first experiment erroneous understanding) of how electrons behave in matter. The theory that $\beta$-rays are monochromatic survived until 1910 and then it was believed that the spectrum consisted of a set of discrete lines due to experiments using the blackening of a photographic plate to record the incident electron velocity spectrum. In the years 1910-13 this method was still very primitive and there was great ignorance about the relationship between the blackening of a photographic plate and the intensity of irradiation. By abandoning the photographic plate method Chadwick discovered the continuous spectrum of $\beta$ decay in 1914. In fact, he did more that this: he showed that photographic plates are extremely sensitive to small changes in the intensity of radiation (so that the importance of the lines is exaggerated) and that the continuous spectrum was superimposed on top of the discrete lines. To find the spectrum Chadwick used a mixture of $Pb^{214}$ and $Bi^{214}$ as a source of "$\beta$-rays" and then measured their intensity by the discharge they caused in an electric potential maintained between a metal plate and a very clean needle with a sharp point.

In the years between 1914 and 1929 there was not much news on the $\beta$-spectrum, however, during this time there was quite a few milestones in physics, some of these were: Einstein writing down the field equations of gravity, Noether finding the theorem relating conservation laws and symmetries, De Broglie introducing wave-particle duality for matter, Pauli conjecturing the exclusion principle, the discovery of spin, Shrödinger writing his first paper on wave mechanics and Dirac conceived the notion of hole theory. On the $\beta$-spectrum front, there was still debate over whether the spectrum was discrete or continuous despite Chadwick's experiment. The debate was finally laid to rest in 1927 by an experiment conceived and performed by Ellis and Wooster. Prior to this experiment it was believed by some that the continuous spectrum was due to other effects such as electromagnetic radiation processes or internal conversion. The idea of the experiment was to look at the energy measured calorimetrically using $\mathrm{Bi}^{210}$ as the source of $\beta$ particles (so that there was no complications with internal conversion). If there was a unique $\beta$ energy which is redistributed by some process then the total energy must be the peak of the $\beta$-spectrum. If, on the other hand, it is only the $\beta$ particle energy that is measured in the calorimeter then the energy per decay would be the average over the spectrum of $\beta$ energies. The experiment revealed the latter! Hence the argument was finally settled and the continuous $\beta$-spectrum won. It is interesting to note, however, that Ellis and Wooster did not venture into interpreting their results and it was quite some time until the seriousness of their result was discussed - that $\beta$ decay seemed to violate energy conservation. Consider the $\beta$ decay process [7],

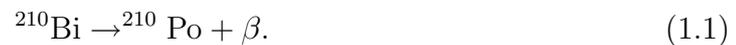

$$^{210}\mathrm{Bi} \rightarrow\, ^{210}\mathrm{Po} + \beta. \tag{1.1}$$

The energy of the $\beta$ particle is the difference between the neutral atomic masses of the initial and final nucleus,

$$E_\beta = m(^{210}\text{Bi}) - m(^{210}\text{Po}) = 1.16 MeV. \tag{1.2}$$

However, recall that for this type of process experiment shows a continuous distribution of energies from 0 to 1.16 MeV. What happened to the missing energy for $E_\beta < 1.16$?

This topic was taken up by Bohr and Pauli in 1929. Bohr advocated that the energy conservation principle should be abolished. Although this may seem shocking in the present, around the time when Bohr had this idea the energy conservation law was questioned on a few occasions. For instance, in 1910 Einstein proposed non-conservation of energy with connection to particle-wave duality (although it was only 3 days later that he realised his mistake) and in 1916 Nernst also suggested, with quantum mechanics as motivation, that conservation of energy is only meaningful in some statistical sense. Pauli on the other hand refused to doubt the energy conservation law ( his defense of this law - and his humour - can be seen in a statement to Bohr, "Do you intend to mistreat the poor energy law further?"). In 1930 Pauli wrote that famous letter proposing that a new particle, which he called the neutron (now known as the neutrino), would save the energy conservation law as $E_\nu + E_e =$ constant ( in the case of the process in Eq.(1.1) $E_\nu + E_e = 1.16 MeV$). He postulated that this new particle will be electrically neutral with spin $\frac{1}{2}$, that it could possibly exist in the nucleus (at this time the real neutron had not been discovered), that it should have a mass of the order of the electron or no more than 0.01 times the proton mass and that its "penetrating power" should be equal to or about ten times larger than the $\gamma$-ray. Recall that at this time only three paricle were known –the proton, the electron and the photon, therefore the proposition of a new particle was bold and courageous.

Even Pauli himself was very cautious and did not announce his idea publicly for a while. By 1936 there was enough evidence to support conservation of energy and Bohr finally sided with Pauli's hypothesis of a new particle. Fermi also adopted Pauli's new particle wholeheartedly and developed a successful theory of $\beta$ radiation using the neutrino in 1934. For a much more thorough review of the conception of the neutrino and references see [8].

## 1.2 Experimental Evidence for the neutrino

The actual experimental observation of the electron neutrino did not come until 1956. The problem was that neutrinos were so weakly interacting that one would need a very large flux of them to be able to detect them. It occured to Reines in 1951, who was working at the Los Alamos Laboratory at the time, that a nuclear bomb explosion would provide the flux needed. Some time later he and Cowen decided to use a fission nuclear reactor instead. The fission products in a nuclear reactor undergo $\beta^-$ decay and consequently emit $\bar{\nu}$. The average emission rate is $6\bar{\nu}$ per fission and the net flux is $10^{13}$ per cm$^2$ per second. The neutrino detector consisted of a proton rich liquid scintillator with a Cd compound. The following reactions take place in the detector in the order written down,

$$
\begin{aligned}
&1.\ \bar{\nu} + p \rightarrow n + e^+ \\
&2.\ e^- + e^+ \rightarrow 2\gamma \\
&3.\ n + {}^{114}Cd \rightarrow {}^{114}Cd^* \rightarrow {}^{114}Cd + \gamma.
\end{aligned}
\tag{1.3}
$$

Hence the detection of the neutrino consisted of the flash of light from electron-positron annihilation and another flash of light approximately $10\mu s$ later due to the capture of the neutron by Cd. Reines and Cowan observed a few events per hour that were candidates for $\bar{\nu}$. However for the evidence to be conclusive many additional experiments were necessary. Finally in 1956, the experiments pointed to the inescapable conclusion that the Pauli's neutrino existed. Reines and Cowan sent a telegram to Pauli giving him this news.

In 1947 the muon was discovered. Soon after this event, the following processes were observed,

$$\pi^+ \to \mu^+ + \nu, \quad \mu^+ \to e^+ + \nu + \bar{\nu}. \tag{1.4}$$

These weak interactions, together with $\beta$ decay, had a common strength. This phenomenon is called the universality of weak interaction. Guided by field theory, it was suggested by many (see [9–11]) that universality could be a result of the interaction of a field with itself by exchange of a charged boson (now known as the W-boson). The existence of the W-boson was put into doubt by Feinberg in 1958. In his paper, [12], Feinberg showed that if a charged boson moderated the weak interactions then the decay $\mu \to e + \gamma$ should occur with a branching ration of $10^{-4}$. However experiments showed that the upper limit was of order $10^{-8}$, showing that a W boson couldn't exist. But the neutrino came along to save the day.

By 1959 it was generally accepted that the neutrinos in pion and moun decay were the same neutrinos that were associated with $\beta$ decay. But it was Lee and Yang who were major influence on resolving the puzzle of the number of neutrino species participating in the reaction $\mu \to e + \gamma$. In 1960 they pointed out at the Rochester Conference that both the W-boson could exist and the process $\mu \to e + \gamma$ be forbidden if two distinct types of neutrinos existed.

Yang and Lee's argument was as follows: the process $\mu \to e + \gamma$ occurs via the intermediate virtual process, $\mu \to e + \nu + \bar{\nu} + \gamma$. If the two neutrinos in this process are of the same type then they would annihilate to nothing, consistent with all physical laws, and the process $\mu \to e + \gamma$ would be allowed to occur. However if the neutrinos were different types then the annihilation would be forbidden and thus the process in question would not be allowed to occur. Together with their argument for the existence of the W boson and the branching ratio if $\nu = \bar{\nu}$ they made a convincing argument for the existence of two types of neutrinos.

Yang and Lee's argument and Feynman's work inspired Schwarz, Steinberger and Lederman to perform the first high energy neutrino experiment to prove the existence of two kinds of neutrinos. The method of the experiment was as follows: make neutrinos via the reaction,

$$\pi^+ \to \mu^+ + \nu \tag{1.5}$$

then detect them using neutrino capture. The possible reactions that could occur in the detector were,

$$\nu + n \;\; \to \;\; p + \mu^- \tag{1.6}$$

$$\nu + n \;\; \to \;\; p + e^- \tag{1.7}$$

If there are two kinds of neutrinos then Eq.(1.6) would occur and Eq.(1.7) would be forbidden. The results of the experiment in 1962 showed that this was indeed the case. The new neutrino was christened as $\nu_\mu$ as it was associated with the muon and the old neutrino was named $\nu_e$ as it was associated with the electron. The results and details of the experiment can be found in [13]. Incidentally, Schwartz, Steinberger and Lederman won a Nobel Prize for the groundbreaking design of the experiment, in particular the method for the high energy neutrino beam, and the discovery of the muon neutrino.

The 20th century was a fantastical journey in the field of particle physics. Technological advances made the implementation of bubble chambers and particle accelerators a reality which in turn introduced mankind to a huge variety of bewildering particles. The conception of the Standard Model explained these particles as combinations of a smaller number of fundamental particles. The Standard Model was also able to logically categorise the fundamental particles. In particular it was postulated that the fermions, consisting of the electron, muon, tau and the neutrinos, formed a doublet structure such that the electron was grouped with $\nu_e$, the muon was grouped with $\nu_\mu$ and the tau was grouped with $\nu_\tau$. The tau lepton was discovered in 1975 suggesting that the tau neutrino should exist.This third type of neutrino, $\nu_\tau$ was only observed directly in 2000 by the DONUT collaboration, [14]. The experiment used a neutrino beam created by $D_S \rightarrow \tau \bar{\nu}_\tau$ and the decay of $\tau$ into another $\nu_\tau$. Some of these tau neutrinos then interacted with an iron nucleus which subsequently produced a tau lepton. The tau lepton decayed leaving a characteristic track. After many months of sifting through the data four tau decay tracks were identified (from a possible 6 million) confirming the existence of the tau neutrino.

## 1.3   The Weak Currents

Recall that the neutrino was invented by Pauli in 1929 and later confirmed to exist by the experiment of Cowan and Reines described in section 1.2. Soon after the conception of the neutrino, in 1934 [15,16], Fermi wrote down the theoretical structure in which to place the new particle. Since the weak interactions were still a "new" phenomenon, Fermi turned to the ideas in quantum electrodynamics to formulate his theory. He reasoned that if an electron could produce a photon so a nucleon could emit an electron and a neutrino. The result of this was the field theoretical form of an interaction involving the neutron, the proton, electron and neutrino fields that would describe the weak interaction. Let us denote the fermion fields by $\psi_i$, then the four-Fermi interaction Hamiltonian density is,

$$H_{weak} = \frac{G_F}{2} \bar{\psi}_p \gamma_\mu \psi_n \bar{\psi}_e \gamma^\mu \psi_{\nu_e} \tag{1.8}$$

where $G_F$ is the Fermi constant.

The difference between the electrodynamic current and the weak current was that during the interaction process the electron retained its charge while a nucleon did not (in $\beta$ decay a neutron changed to a proton). Subsequently processes without change of charge were termed "neutral" while those with such a change were called "charged". In the next 30 years, a wealth of experimental data became available which necessitated the modification of the Fermi's interaction Hamiltonian density. It was also assumed that weak currents were only charged since these experiments indicated that neutral processes were suppressed if they were present at all. By the early 1970's Fermi's "point interaction" was cast into doubt. Unlike the electrodynamics, Fermi's theory was not renormalizable. The solution to this problem was to invoke an intermediate vector boson, W, again with the analogy to electrodynamics. It was thought that the W boson would be charged due to the assumed non-existence of the neutral weak current. Hence it became fashionable to search for the W boson experimentally. One of these bubble chamber experiments, Gargamelle at CERN, led to the discovery of the neutral current.

The search for the neutral current only become feasible when theoretical motivations came to light through the early conception of the Standard Model. The Yang and Mills theory of weak and electromagnetic interaction, showed that through the spontaneous breaking of symmetry there is a natural relationship between the weak interactions and electromagnetism. Weinberg [17] and Salam [18] showed that a new neutral current must exist in order for the theory to be complete. However, their theory was not taken seriously until 't Hooft showed that it was renormalizable [19].

Not long after the appearance of 't Hooft's paper Zumino, Prentki and Gaillard convinced the Gargamelle group at CERN to take up the challenge of finding the neutral current experimentally. It turned out that the first experimental proof occurred in January 1973 due to an accidental find of a single process of the kind $e + \nu_\mu \rightarrow e + \nu_\mu$, which is only allowed to occur via the neutral current. Over the following years this result was corroborated in a number of experiments. The complete review on the experiments that led to the confirmation of neutral current can be found in [20] and [21].

Today, a great deal more is known about the Lorentz structure of the interaction and many more weak processes are known to be part of $H_{weak}$. As a consequence the four-Fermi interaction has been modified,

- The space time structure of the currents has both the vector currents as in Eq.(1.8) and axial currents.

- The weak interactions do not consist only of charged current processes but also neutral current processes.

- The fundamental fields in the Hamiltonian density are fields describing the quarks rather than the hadronic fields such as the proton and neutron.

With these changes the $H_{weak}$ can be rewritten in a more compact form that includes both the neutral current $J^\mu$ and charged current $K^\mu$,

$$H_{weak} = \frac{4G_F}{\sqrt{2}}[J^\mu(x)J_\mu^\dagger(x) + \rho K^\mu(x)K_\mu(x)]. \qquad (1.9)$$

## 1.4 Modern Neutrino Physics

The humble neutrino has left an important imprint on particle physics in the past few decades and continues to make an impact. Due to the observation in the 1960's that the number of neutrinos arriving from the sun was less than that predicted by the Standard Solar Model the theory of neutrino flavour change through the phenomenon of neutrino oscillations was conceived. Through pioneering experiments such as Kam-Land, SNO, SuperKamiokande, Chooz and many more, neutrino oscillations have been confirmed. One of these experiments, LSND, stands out like a sore thumb indicating that there may be another type of neutrino, named the sterile neutrino, which does not interact with matter at all but which the active neutrinos ($\nu_e$, $\nu_\mu$, $\nu_\tau$) may oscillate into. However, as time goes by, other experiments limit this possibility severely. The pioneering theories of Wolfenstein, Smirvov and Michayev showed that neutrinos can change flavour due to the charged current and neutral current interaction with the medium. The MSW theory has also contributed to the solution of the solar neutrino problem.

Although neutrino oscillations is a phenomenal discovery on its own it is its implications for particle physics which are very important. The fact that the neutrinos are not massless and that there may exist a sterile neutrino is an indicator that the Standard Model of particle physics is not the whole story. It is also interesting that the neutrino mass and coupling constant are on scale many orders of magnitude smaller than that of the other fundamental particles. At the same time it is these properties of the neutrino that allows one to propose theoretical solutions to many problems in cosmology. For example, their elusive nature and their abundance has been used to propose it as a candidate for dark matter, the nature of their interactions may lead to explaining the baryon-lepton asymmetry, their non-interacting property may lead to explaining the supernova explosion. The neutrino is also an important ingredient in generating the light elements in the early universe (nucleosynthesis) thus explaining the helium abundance in the universe and also in generating heavy elements in supernovaee.

Just as understanding neutrinos interacting with matter has been an important study, the interaction of neutrinos with themselves is also essential to understanding many systems in the universe. In systems such as the early universe and supernovae the neutrino number density is far greater than or of the same order as the electron and baryon density so one must consider neutrino self-refraction. Neutrino self-interactions have been a tougher problem to solve than than the interaction of neutrinos with matter such as electrons. Naturally, some physicists, [5, 6], were inclined to treat neutrino-neutrino interactions analogously to the MSW theory. However McKellar and Thomson [22], Sigl and Raffelt [23], and Pantaleone [1, 2], showed that this was not quite the right way to go about it. These physicists showed flavour could be exchanged between the neutrinos through the interaction which is not predicted by the MSW theory since the neutrinos only interact via the neutral current. This theory has been used in constructing equations of motion in the early universe and supernovae. However as it is an understandably tough task to construct and analyse these equations many assumptions need to be made. This thesis is concerned with checking whether one of these assumptions is correct. The assumption we will be focused on is that the wave function of the interacting neutrinos can be factorised into one body states. In other words, we ask the question, do the neutrinos become quantumechanically entangled and, if so, what effect does this have on the neutrino ensemble?

# Chapter 2

# Neutrino Flavour Conversion

In the 1960's, an experiment at the Homestake mine detected electron neutrinos coming from the sun. The experiment found a large discrepancy between the number of electron neutrinos detected and the number of electron neutrinos predicted by the standard solar model [24]. In the decades that followed two theories were put forward to explain the discrepancy that have stood the test of time: neutrino oscillations and the MSW theory of neutrinos in matter. Both of these theories propose the flavour change of the neutrino which renders it invisible to the detector rather than the "disappearance" of the neutrino.

In later years scientists began to look outside of our solar system and began to study neutrinos in the early universe and the interior of stars, particularly supernovae. The pioneering theories of neutrino oscillations and MSW made their presence known in these systems. In these environments an extension to the MSW theory was also required to encompass neutrino-neutrino interactions as the number density of the neutrinos is a sizable fraction of the total particle number density in the early universe and supernovae.

In this chapter we review these theories so that the reader is able to have an understanding for the motivation of the content in this thesis. In section 2.4 we point out some of the recent work in the area of flavour conversion in a neutrino ensemble which questions the validity of the models constructed for the early universe and supernovae. Finally we describe generally the work we have contributed in this area which the rest of this thesis will explain in detail.

## 2.1   Neutrino Flavour Conversion in the Vacuum

In the Standard Model each of the charged leptons e, $\mu$ and $\tau$ is coupled to a neutrino $\nu_e$, $\nu_\mu$ and $\nu_\tau$ respectively. The simplest form of the Standard Model takes the neutrino mass as zero, however the evidence for neutrinos to be massive is now overwhelming. The usual analysis of neutrino mixing is a nice example of quantum mechanics. The flavour neutrinos do not have definite mass, rather they are superpositions of mass eigenstates with distinct mass. We begin with a short review of the flavour eigenstates and their relation to the mass eigenstates.

The flavour neutrino is a superposition of mass eigenstates. Considering only two neutrino species ($\nu_e$, $\nu_\mu$) for simplicity, the flavour neutrino state is,

$$| \, \nu_e \, \rangle \;\; = \;\; \cos\theta | \, \nu_1 \, \rangle + \sin\theta | \, \nu_2 \, \rangle \qquad (2.1)$$

$$| \, \nu_\mu \, \rangle \;\; = \;\; \cos\theta | \, \nu_2 \, \rangle - \sin\theta | \, \nu_1 \, \rangle. \qquad (2.2)$$

A pictorial representation can be found in figure 2.1(a) . Here $\nu_1$ and $\nu_2$ are the mass eigenstates. They are the eigenstates which diagonalise the Hamiltonian.The mixing angle ,$\theta$, is now known to be almost maximal. The current best fit is $\theta \approx 34°$ [25].

Equivalently we can write down the mass eigenstate ($\nu_1$, $\nu_2$) in terms of the flavour states,

$$| \nu_1 \rangle = \cos\theta | \nu_e \rangle - \sin\theta | \nu_\mu \rangle \qquad (2.3)$$

$$| \nu_2 \rangle = \cos\theta | \nu_\mu \rangle - \sin\theta | \nu_e \rangle. \qquad (2.4)$$

In this form, the states describe the flavour content of the mass states. Hence, for example, if the neutrino is born as a $\nu_1$ the probability that it will be detected as an electron neutrino is $\sin^2\theta$. Figure 2.1(b) shows this pictorially.

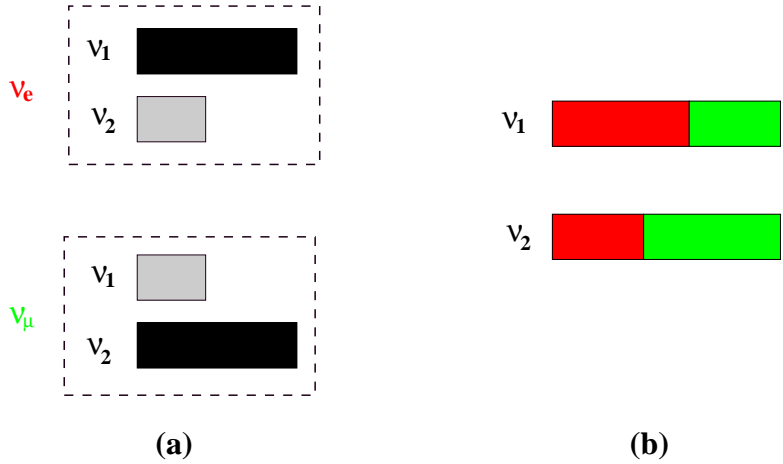

**(a)**                    **(b)**

Figure 2.1: (a) Representation of flavour neutrinos in terms of mass eigenstates. The length of the boxes represents the probability of finding a mass eigenstate in a given flavour state. (b) Representation of mass eigenstates in terms of the flavour content. The colour red represents the electron neutrino while the colour green represents the muon neutrino. The size of the red or green box indicates the probability of finding the electron neutrino or the muon neutrino in a given mass eigenstate

The time evolution of the neutrino is determined by the Hamiltonian in quantum mechanics. The vacuum Hamiltonian in the mass basis is,

$$H_{vac}^{mass} = \begin{bmatrix} E_1 & 0 \\ 0 & E_2 \end{bmatrix} \approx \begin{bmatrix} E & 0 \\ 0 & E \end{bmatrix} + \frac{1}{2E} \begin{bmatrix} m_1^2 & 0 \\ 0 & m_2^2 \end{bmatrix}, \quad (2.5)$$

where we have assumed that the masses of the mass eigenstates are small ands used the approximation $E = \sqrt{m^2 + p^2} \approx E + \frac{m^2}{2p}$ and $|p| \approx E$. In general the term proportional to the identity matrix can be dropped as the it does not contribute to neutrino oscillations.

We can rotate the Hamiltonian in the mass basis to the flavour basis with the aid of the mixing matrix,

$$U = \begin{bmatrix} \cos\theta & \sin\theta \\ -\sin\theta & \cos\theta \end{bmatrix}. \quad (2.6)$$

Ignoring a term proportional to the identity matrix, the Hamiltonian in the flavour basis is,

$$H_{vac}^{flavour} = \frac{1}{2E} U \begin{bmatrix} m_1^2 & 0 \\ 0 & m_2^2 \end{bmatrix} U^\dagger \quad (2.7)$$

$$= \frac{\Delta m^2}{4E} \begin{bmatrix} -\cos 2\theta & \sin 2\theta \\ \sin 2\theta & \cos 2\theta \end{bmatrix}, \quad (2.8)$$

where $\Delta m^2 = m_2^2 - m_1^2$.

The eigenvalues of the Hamiltonian are,

$$\lambda_1 = \frac{m_1^2}{2E} \qquad \lambda_2 = \frac{m_2^2}{2E}. \quad (2.9)$$

According to quantum mechanical time evolution a neutrino which is born at $t = 0$ as an electron neutrino will evolve in time as,

$$| \nu(t) \rangle = e^{\frac{im_1^2}{2E}t} \cos\theta | \nu_1 \rangle + e^{\frac{im_2^2}{2E}t} \sin\theta | \nu_2 \rangle. \tag{2.10}$$

Note that an overall irrelevant phase has been emitted due to discarding terms proportional to the identity matrix in the Hamiltonian. Since the phases change with time, the neutrino which started out as an electron neutrino could change into a muon neutrino. The probability of finding a muon neutrino some time later is,

$$P(\nu_e \to \nu_\mu) = |\langle \nu(t) \mid \nu_\mu \rangle|^2 = \sin^2\theta \sin^2\left(\frac{\Delta m^2}{4E}t\right), \tag{2.11}$$

Notice that the probability is oscillatory and that there must be a difference between the mass of the mass eigenstates in order for the flavour conversion to take place. For further discussion see [26] and [27].

The neutrino oscillation theory, founded by Pontecorvo [28] in 1958, was a candidate for solving the electron neutrino deficit in the Homestake experiment. The theory shows that the electron neutrinos have changed into muon neutrinos while propagating in the vacuum. The phenomenon of neutrino oscillations is an accepted theory today but it is only part of the solution to the solar neutrino problem.

## 2.2   Neutrino Flavour Conversion in Matter

In the early days it was thought that neutrino oscillations in vacuum could not explain the discrepancy by themselves as it was hypothesised that the mixing angle, which determines the oscillation amplitude, was small. This belief was motivated by the observation that the mixing angle was small in the hadronic sector. Hence some attention was given to the study of neutrino propagation in the sun since solar neutrinos are born in the centre of the sun and travel in matter to the edge of the sun. Wolfenstein [29] and Mikheyev and Smirnov [30–32] showed that the interactions of neutrinos with matter could produce a large effect.

The essence of their theory is that neutrinos in a medium gain an effective mass due to the interactions with matter particles. The electron neutrinos interact differently with normal matter (electrons, protons, neutrons) than other neutrinos, hence they have a different effective mass to the other neutrinos. The difference in the effective mass can lead to a change of flavour of the neutrino. This is analogous to vacuum oscillations in the sense that flavour change in the vacuum is driven by the difference in the mass of the mass eigenstates. Of course one must consider both the vacuum effect plus the matter effect when analysing neutrino propagation in matter. However the matter effect can dominate in some situations. Here we will discuss some of the details of this theory so that we may draw parallels when discussing neutrino self-interactions.

In the literature the matter is called the background while the propagating neutrinos is called the beam. At low energies only the elastic forward (coherent) scattering is relevant and the inelastic scattering can be ignored. Assuming that neutrino interactions with matter conserve the neutrino flavour, as described in the Standard Model, the coherent scattering of neutrinos occur in two ways for electron neutrinos. The electron neutrinos can interact with electrons via W exchange *and* via Z exchange. Other neutrinos can only interact with the medium via Z exchange. The neutral current is flavour conserving and affects all neutrinos equally. The Feynman diagrams for the interactions of neutrinos with matter are depicted in figure 2.2.

Coherent scattering of electron neutrinos via the charged current gives rise to an interaction potential energy $V_W$ which clearly depends on the number density of electrons, $\rho_e$,

$$V_W = \sqrt{2} G_F \rho_e. \tag{2.12}$$

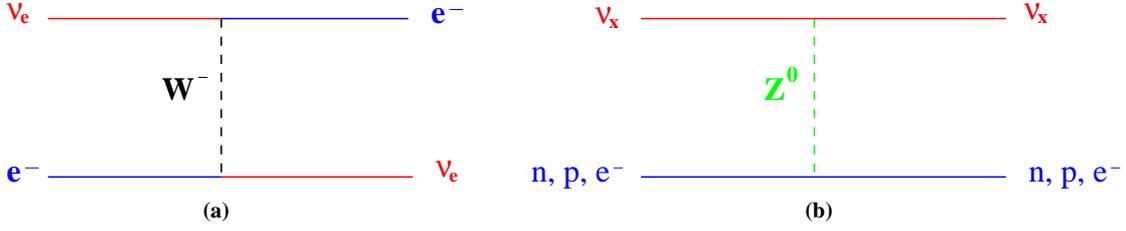

Figure 2.2: Feynman diagrams for neutrinos interacting with matter. (a) The electron neutrino interacts with the electron via the charged current. (b) All flavour neutrinos interact with matter via the neutral current. Note that $x = e, \mu, \tau$.

The coherent scattering of neutrinos via the neutral current only depends on the interaction with the neutron if the medium is electrically neutral. The Z coupling of the neutrino to the electron and proton is equal and opposite, hence if there are equal numbers of electrons and protons in the medium, their contribution to the coherent forward neutrino scattering potential cancel out. Thus the Z exchange gives rise to a flavour independent potential which is proportional to the number density of the neutrons,

$$V_Z = -\frac{\sqrt{2}}{2} G_F N_n. \tag{2.13}$$

In the sun, to a good approximation, only two neutrinos, $\nu_e$ and $\nu_\mu$, play a significant role. Therefore the interaction Hamiltonian in the flavour basis is:

$$H_{mat} = V_W \begin{bmatrix} 1 & 0 \\ 0 & 0 \end{bmatrix} + V_Z \begin{bmatrix} 1 & 0 \\ 0 & 1 \end{bmatrix}. \tag{2.14}$$

Since the neutral current interaction with the medium is the same for all neutrinos it is proportional to the identity matrix and can be dropped. Further, it has been confirmed that the masses of the mass eigenstates ($\nu_1$, $\nu_2$, $\nu_3$) are not equal. A mass difference between the mass eigenstate leads to neutrino oscillations hence we must add this term to the Hamiltonian. The total Hamiltonian in matter is,

$$
\begin{aligned}
H &= H_{vac} + H_{mat} \\
&= \frac{1}{2E} U \begin{bmatrix} m_1^2 & 0 \\ 0 & m_2^2 \end{bmatrix} U^\dagger + V_W \begin{bmatrix} 1 & 0 \\ 0 & 0 \end{bmatrix}
\end{aligned}
\tag{2.15}
$$

.

Notice that the total Hamiltonian is no longer diagonal in the vacuum mass basis. However it is possible to find a new basis for which the Hamiltonian is diagonal. This is the matter mass basis with eigenstates $\nu_{1m}$ and $\nu_{2m}$. The matter mass basis can be found by defining the matter mixing angle, $\theta_M$ and the matter mass splitting (or the difference between the square effective masses of the matter mass eigenstates), $\Delta m_M$. For a medium with constant density,

$$
\Delta m_M^2 = \Delta m^2 \sqrt{\sin^2 2\theta + (\cos 2\theta - A)^2}
\tag{2.16}
$$

$$
\sin^2 2\theta_M = \frac{\sin^2 2\theta}{\sin^2 2\theta + (\cos 2\theta - A)^2}.
\tag{2.17}
$$

Here $A = \frac{4EV_W}{2\Delta m^2} = \frac{2\sqrt{2}G_F n_e E}{\Delta m^2}$. It is clear that the matter mass splitting of $\Delta m_M^2$, Eq.(2.16) is the effective mass difference due to the different interaction of the electron neutrino with the electron compared to the other neutrinos since it contains $V_W$. The

Hamiltonian in the flavour basis can now be written as,

$$H = \frac{\Delta m_M^2}{4E} \left[ \begin{array}{cc} -\cos 2\theta_M & \sin 2\theta_M \\ \sin 2\theta_M & \cos 2\theta_M \end{array} \right].$$ (2.18)

This is analogous to the vacuum Hamiltonian Eq(2.8). Hence one can proceed as for the vacuum and find neutrino oscillations depend on the matter mixing angle , $\theta_M$, and the matter mass splitting, $\Delta m_M^2$. These can differ greatly from the vacuum mixing angle, $\theta$, and the vacuum mass spllitting, $\Delta m^2$, depending on the density of the medium and the energy of the neutrino, hence leading to very different probabilities.

However the density in the sun is not constant. This can produce flavour change in a different way to neutrino oscillations. In a non-constant medium the matter mixing angle and the effective mass squared difference depends on the denisty (or equivalently it changes in time as the neutrino propagates into a region of different density). We can write down the matter mass eigenstates in terms of the flavour states,

$$| \nu_{1m} \rangle = \cos\theta_M(\rho_e)| \nu_e \rangle - \sin\theta_M(\rho_e)| \nu_\mu \rangle$$ (2.19)

$$| \nu_{2m} \rangle = \cos\theta_M(\rho_e)| \nu_\mu \rangle - \sin\theta_M(\rho_e)| \nu_e \rangle.$$ (2.20)

Therefore, the flavour content of a given mass eigenstate depends on the density of the medium.

High energy neutrinos, produced via $^8B$, are born in the dense region in the centre of the sun. The Hamiltonian for this is system is that of Eq.(2.15) . Initially, because of the high energy of the neutrinos and the density of electrons, $H_{mat}$ dominates and as a first approximation (which is sufficient for this discussion) we can ignore $H_{vac}$. The Hamiltonian is now diagonal in the flavour basis and the electron neutrino is the higher energy eigenstate of the Hamiltonian, $\nu_{2m}$, with effective mass $V_W$.

The matter mass eigenstate $\nu_{1m}$ with zero effective mass is non-existent. Hence, here, neutrino oscillations is suppressed.

The propagation of the neutrino from the centre to the edge of the sun is adiabatic. That is, the density of matter decreases slowly with distance and so the contribution of the vacuum masses can no longer be ignored. The neutrino will propagate outwards as the slowly changing eigenstate of the slowly changing Hamiltonian. Approximately half way through this journey the matter mass eigenstate, $\nu_{2m}$, becomes an equal mixture of an electron neutrino and a muon neutrino . At the edge of the sun, $\nu_{2m}$ is approximately 2/3 muon neutrino and 1/3 electron neutrino. The journey of the neutrino is depicted in figure 2.3. In this figure the bar represents $\nu_{2m}$, the colour red represents the content of the electron neutrino and the colour green represents the content of the muon neutrino.

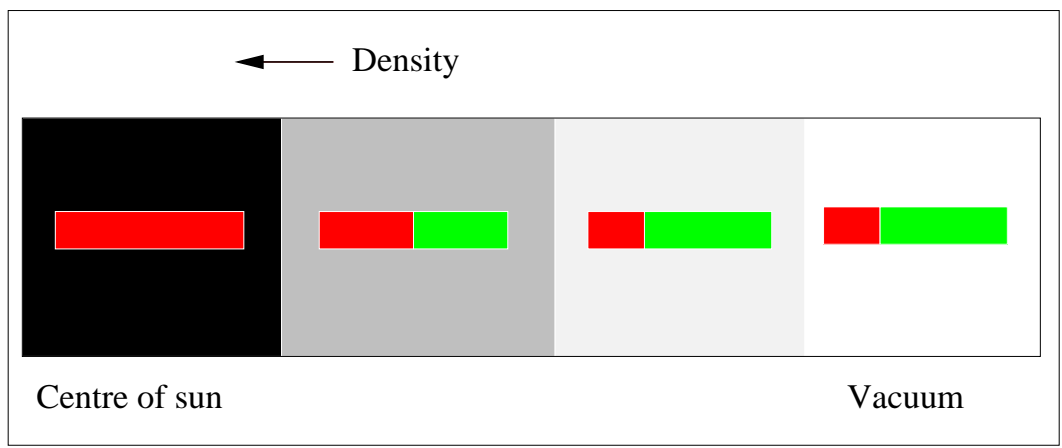

Figure 2.3: A representation of a neutrino undergoing adiabatic conversion in the sun. The neutrino begins its journey from the centre of the sun. The rectangular box indicates the matter mass eigenstate, $\nu_{2m}$. The red colour represents the content of the electron neutrino and the green colour represents the content of the muon neutrino. Hence the neutrino begins as an electron neutrino and as it enters the vacuum it is a mixture of an electron and muon neutrino.

At the boundary between the vacuum and edge of the sun, the matter mass eigenstate is equal to the vacuum mass eigenstate i.e. $\nu_{2m} = \nu_2$. The probability of finding an electron neutrino as it emerges into the vacuum is non-oscillatory,

$$P = |\langle\,\nu_e\mid\nu(t)\,\rangle|^2 \approx |\langle\,\nu_e\mid\nu_{2m}(t)\,\rangle|^2 = |\langle\,\nu_e\mid\nu_2(t)\,\rangle|^2 \approx \sin^2\theta \qquad (2.21)$$

.

This is an example of flavour change due to the interaction with the medium only. It also illustrates the potential for the medium to have a huge impact on the flavour evolution of the neutrino and neutrino oscillations. For further reading see [27] and [33].

## 2.3 Neutrino Flavour Conversion via Neutrino Self-Interactions

In situations like the Sun and the Earth neutrino self-interactions are ignored as the neutrinos are weakly interacting and the number density of normal matter ($n_m$) greatly exceeds the number density of neutrinos ($n_\nu$). However there are physical systems where one needs to take into account neutrino-neutrino interactions.

In the early universe neutrino self-interactions are important since the number denisty of neutrinos is approximately the same as that of electrons and positrons before $e^+e^-$ annihilation and much greater than the number density of baryons and electrons (positrons) after $e^+e^-$ annihilation. Neutrinos in this era of the universe contribute to the lepton chemical potential, they dictate the abundance of light elements (big bang nucleosynthesis) which in turn puts an upper limit on the number of lepton generations. See [34] for a full description.

Similarly, near the core of a supernova the number density of neutrinos is approximately the same as that of baryons. More than 95% of the gravitational binding energy released when the core of a massive supernova collapses to a hot neutron star is trapped in the collapsing core in the form of a degenerate sea of electron type neutrinos. Tapping into the energy reservoir of the degenerate electron neutrino sea could change the energetics of the supernova collapse. In the supernova, neutrinos also contribute to the making of heavy elements (synthesis of heavy elements). See [35] for an introduction to neutrino physics in supernovae.

In these dense neutrino systems the flavour evolution is linked to the approach or departure from equilibrium. Hence it is important to track the flavour of the neutrino as it coherently scatters off other neutrinos. Notzold and Raffelt [6], and Fuller et. al [5] recognized that neutrino self-interactions must play a significant role in the early universe and supernova. They treated the problem by analogy with MSW. Recall that the difference in the induced mass squared plays a significant part in flavour conversion in MSW theory. The induced mass is dependant on neutrino interactions. Fuller et. at. [5] were interested in amplification of lepton number violation through the interaction of the electron neutrino with another heavier neutrino denoted $\nu_x$ (such as a $\nu_\tau$ or $\nu_\mu$). The interactions they considered are shown in figure 2.4.

Notzold and Raffelt [6] were interested in magnetically induced left-right oscillations in the early universe. However their results also encompass the interior of stars as the conditions here resemble that of the early universe. For their analysis it was important to calculate the induced mass squared differences due to all neutrino interactions ($\nu - \nu$ interactions $e - \nu$ interactions, nucleon$- \nu$ interactions). The $\nu - \nu$ coherent interactions are depicted in figure 2.5, while the rest of the interactions are as in figure 2.2 .

The common link between the interactions in figures2.4 and 2.5 is that the induced mass terms are diagonal in the flavour basis. Recall that this was also the case for neutrinos in the sun. Hence one can then proceed as with the case for normal matter and construct a matter mass basis for which the Hamiltonian is diagonal and track the flavour evolution of the neutrino.

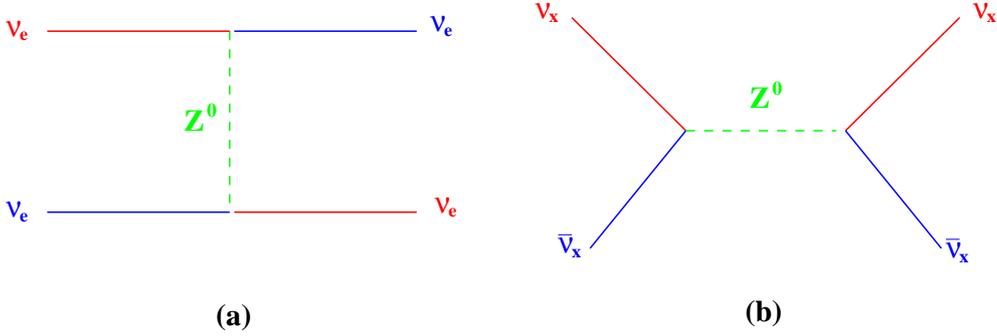

Figure 2.4: Feynman diagrams for neutrino-neutrino scattering in the analysis of supernovae by Fuller et.al. [5]. (a) The neutral current exchange contribution to for $\nu_e - \nu_e$ scattering. (b) The annihilation contribution for $\nu_x - \bar{\nu}_x$ scattering. A similar diagram was also considered for $\nu_e - \bar{\nu}_e$ scattering.

The neutral current weak interaction Hamiltonian is U(k) invariant (k=number of neutrino species):

$$H_{NC} = \frac{G_F}{\sqrt{2}} \left( \sum_{\alpha} j_{\alpha}^{\mu} \right) \left( \sum_{\beta} j_{\beta\mu} \right) \tag{2.22}$$

.

Hence the interaction Hamiltonian constructed for neutrinos in supernova and early universe should also respect this symmetry. In order for the Hamiltonian to be invariant under U(k) it must be non-diagonal in the flavour basis. While Notzold and Raffelt, and Fuller correctly calculated the diagonal contributions they were obviously missing the off diagonal components. The pioneers who derived the off-diagonal contribution to the neutrino-neutrino interaction Hamiltonian were McKellar and Thomson [22], Pantaleone [1,2] and Sigl and Raffelt [23]

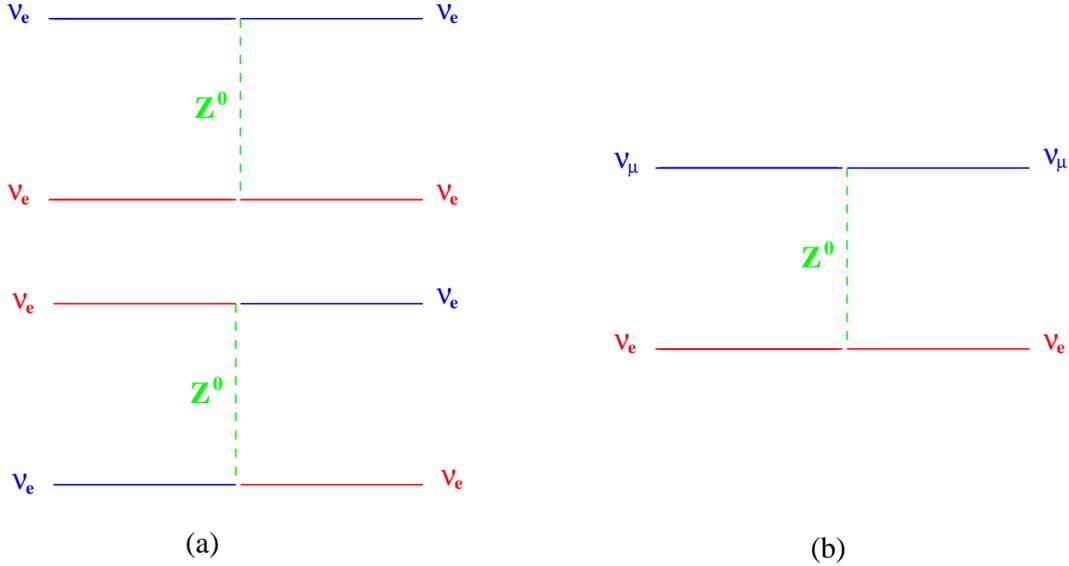

Figure 2.5: Notzold and Raffelt [6] neutrino-neutrino interactions in the early universe and supernova. These diagrams represent an electron neutrino (red) interactions with other neutrinos. Similar diagrams exist for other types of neutrinos. (a) $\nu_e - \nu_e$ interactions. (b) $\nu_e - \nu_\mu$ interactions. A similar diagrams exists for $\nu_\tau$.

In analysing the neutrino ensemble it is convenient to talk about the system in terms of the beam and the background. Here, one is generally interested in the flavour evolution in the beam. With this definition, the flavour off-diagonal terms can be interpreted as an exchange of flavour between the beam neutrino and the background neutrino or equivalently as the exchange of the momentum between the neutrinos. Figure 2.6 shows the diagonal and off-diagonal contributions to the interaction Hamiltonian.

It is important to note, however, that total flavour is conserved but the flavour associated with each neutrinos is not conserved. To conserve total flavour, the background neutrino must be scattered into the beam or vice-versa. This is shown pictorially in figure 2.7.

The importance of the off-diagonal terms stems from the exchange of different momenta between the neutrinos which may cause entanglement and thus any model constructed for a dense neutrino system is in general a many-body problem. To illustrate this we write down the equation for the flavour evolution of interacting neutrinos. Consider two neutrinos described by plane waves with momentum $\vec{k}$ and $\vec{q}$. There are four possibilities with two flavours, leading to the equation [2, 4],

$$i\frac{d}{dt}\begin{pmatrix} |\nu_e(\vec{k})\nu_e(\vec{q})\rangle \\ |\nu_e(\vec{k})\nu_\mu(\vec{q})\rangle \\ |\nu_\mu(\vec{k})\nu_e(\vec{q})\rangle \\ |\nu_\mu(\vec{k})\nu_\mu(\vec{q})\rangle \end{pmatrix} = V_{\nu\nu}\begin{pmatrix} |\nu_e(\vec{k})\nu_e(\vec{q})\rangle \\ |\nu_e(\vec{k})\nu_\mu(\vec{q})\rangle \\ |\nu_\mu(\vec{k})\nu_e(\vec{q})\rangle \\ |\nu_\mu(\vec{k})\nu_\mu(\vec{q})\rangle \end{pmatrix} \qquad (2.23)$$

where the interaction potential, $V_{\nu\nu}$, is found by considering all the possible interactions,

$$V_{\nu\nu} = \frac{\sqrt{2}G_F}{V}(1 - \cos\beta)\begin{pmatrix} 2 & 0 & 0 & 0 \\ 0 & 1 & 1 & 0 \\ 0 & 1 & 1 & 0 \\ 0 & 0 & 0 & 2 \end{pmatrix}. \qquad (2.24)$$

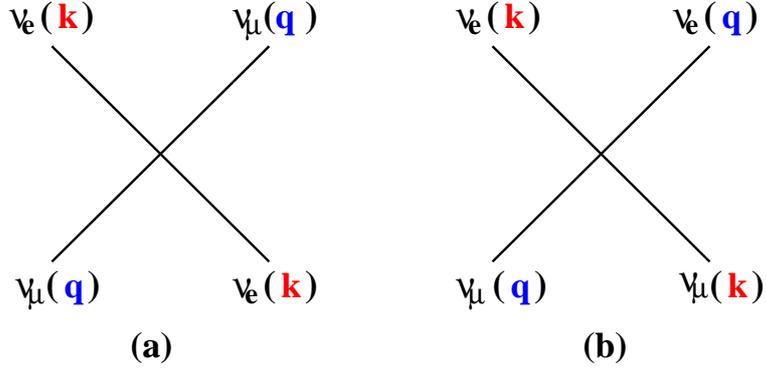

Figure 2.6: The (a) diagonal and (b) off-diagonal contributions to the interaction Hamiltonian.

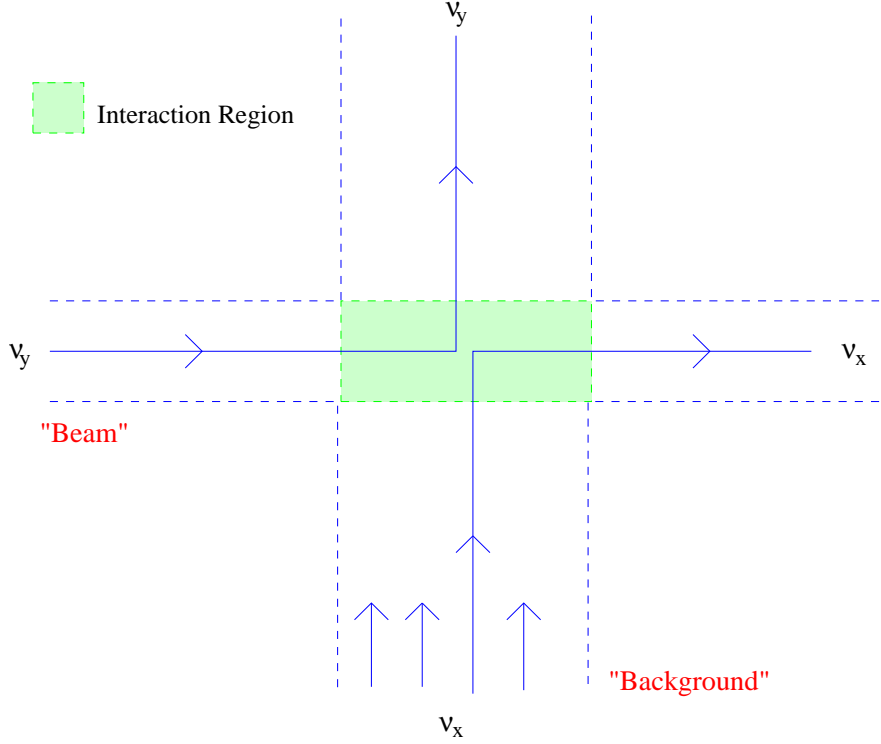

Figure 2.7: The background neutrino $\nu_x$ scatters into the beam due to the exchange of momenta between the beam neutrino $\nu_y$ and background neutrino $\nu_x$.

The angle $\beta$ is the angle between the two momenta. Now consider a test electron neutrino interacting with two muon neutrinos in the box. The two scattering events are described by Eq.(2.23). Keeping only the forward scattering with angle $\beta = \Theta$, the result of the interaction, using a small t expansion, is [4],

$$|\nu_e\rangle|\nu_\mu\nu_\mu\rangle \longrightarrow (1 + 2iA)|\nu_e\rangle|\nu_\mu\nu_\mu\rangle + iA(|\nu_\mu\rangle|\nu_e\nu_\mu\rangle + |\nu_\mu\rangle|\nu_\mu\nu_e\rangle) \qquad (2.25)$$

where $A = -\frac{\sqrt{2}G_F t}{V}(1 - \cos\Theta)$. The second term in the equation shows that the electron neutrino has been exchanged with the muon neutrino. Furthermore, Eq.(2.25) shows that each neutrino-neutrino interaction produces an entangled state.

## 2.4   The One-Body Description of a Neutrino Ensemble

McKellar and Thomson [22],Pantaleone [1,2] and Sigl and Raffelt [23] derived the off-diagonal terms in the course of constructing equations of motion for the early universe and supernova. These authors constructed the equation of motion of the neutrino system by analogy with neutrinos in a background of electrons, by splitting the system into beam and background (although in the case of $\nu - \nu$ interactions the distinction is arbitrary) and thus ending up with a one body equation. This approximation is valid if the state of the system can be factorized into a product of one particle states. If a system is entangled or has dynamical corrrelations then it is not factorizable into single-body states. It is possible for neutrinos to develop entanglement due to the off-diagonal induced mass terms. So the question arises whether this entanglement has a substantial impact on the flavour evolution of the neutrino ensemble. If many-body neutrino interactions are important it could have significant impact on the conclusions that can be drawn from previous models.

The question was recently examined by Friedland and Lunardini [4,36] and by Bell et. al. [3] . All three papers used in principle a similar setup, with a Hamiltonian that was restricted to forward scattering only in which the neutrinos scatter coherently. While Friedland and Lunardini [36] argued that the coherent part of the neutrino evolution should be described by the one-particle formalism, Bell et. al. reached an opposite conclusion.

Bell et. al. considered the evolution of a system initially in the flavor eigenstates. With this choice, the one-particle formalism predicts no coherent flavor conversion and thus conversion on the coherent time scale in this system would be an indication of the breakdown of the one-particle description (presumably through the formation of many-neutrino entangled states). The numerical calculation performed by Bell et. al. seemed to suggest the presence of such fast conversions, although the calculations involved a relatively small numbers of neutrinos. Subsequently, reference [4] solves the neutrino model introduced by Bell *et. al.* analytically for the special case of equal numbers of each neutrino species and equal strength interactions by mapping neutrino-neutrino interactions to spin-spin interactions. They found in the limit of infinitely many particles the time to reach flavour equilibrium is of an incoherent nature thus supporting the one-particle description of the system.

This thesis is concerned with understanding the many-body neutrino model further. In particular we will extend the work of Friedland and Lunardini by considering a system of many neutrinos in which there are only two species and generalise the model to initial states in which the species are not equally populated. Our conclusion is that the one particle description of the system is valid. Further we find that the quantum system exhibits a rich set of behaviours and interesting physics.

The problem of the flavor evolution in dense neutrino systems continues to receive a significant amount of attention. In addition to the above mentioned papers [3, 4, 36], other notable references include [37–43].

# Chapter 3

# Construction of a Simplified Many-Body Neutrino Model

We are interested in investigating coherent effects in a neutrino system. As explained in the previous chapter it is the coherent scattering of the neutrinos which may cause quantum correlations or entanglement, which, in turn, can cause a breakdown of the one-body approximations employed by physicists studying the flavour evolution of neutrinos in the early universe and supernovae. Coherent scattering of neutrinos occurs when the neutrinos preserve momenta or exchange momenta. In the literature this is called forward scattering. To study the effects of entanglement and whether it plays a significant role in the flavour evolution of the neutrino, we construct a density matrix of the many-neutrino system. The analysis consists of constructing the probability that a neutrino initially in a particular flavour eigenstate (such as an electron neutrino) remains in the same flavour eigenstate. This allows us to track the flavour evolution of this neutrino and hence establish the importance of many-neutrino entanglement.

In this chapter we present a generalisation of the many-body neutrino model intro-duced by Friedland and Lunardini (F&L) [4]. The F&L model considered a system of many neutrinos in which there are two flavour species, where the number of neutrinos of one flavour was equal to the number of neutrinos of the other flavour. For example, N electron neutrinos and N muon neutrinos. Further their model considered only neutrinos which were in flavour eigenstates. We generalise the F&L model to initial states in which the species are not equally populated. The model is constructed analytically by comparing the two neutrino system to a spin system. The model has the following features which were also present in the F&L system as well as the system studied by Bell *et. al.*,

- We represent the neutrino system as plane waves in a box with volume $V$.

- We consider only massless neutrinos hence we do not include neutrino oscillations. The advantage of this is that our results will clearly be related only to the flavour evolution of neutrinos due to $\nu - \nu$ coherernt scattering. However we discuss including neutrino oscillations into the picture in chapter 6.

- We are interested in investigating coherent neutrino scattering hence the model only keeps the forward scattering terms. The model should correctly capture coherent effects in a real neutrino system but not necessarily all the incoherent scattering effects.

- The momentum degrees of freedom are ignored. Hence the effects of Fermi statistics are not included.

- The interaction strength between any two neutrinos is taken to be constant so that the model can be solved analytically. This means that we are ignoring the angular distribution of the neutrino momenta. The model is thus relevant for an isotropic neutrino but may not capture all the effects that could arise due to anisotropic momentum distributions. Sawyer [42, 43] has considered the case of anisotropic momentrum distributions.

- The solution and subsequently any analysis of the many-body system is valid for any two state system with constant couplings.

## 3.1  Setup: The Hamiltonian and Eigenvalues

For $E_\nu << m_{Z^0}$ the neutral current interaction Hamiltonian is,

$$H_{int} = \frac{G_F}{2} \left( \sum_\alpha \bar{\nu}_\alpha \gamma^\mu \nu_\alpha \right) \left( \sum_\beta \bar{\nu}_\beta \gamma^\mu \nu_\beta \right). \tag{3.1}$$

Here the sum is over all flavours. Let us make the simplifying assumption that there are only two neutrino species. The Hamiltonian is then invariant under SU(2). Therefore $\nu - \nu$ scattering is equivalent to spin-spin coupling as explicitly shown in [4]. We present this argument for completeness. Consider two interacting neutrinos described by the flavour conserving neutral current Hamiltonian of Eq.(3.1). According to this Hamiltonian, the flavour space wavefunction of an outgoing neutrino is equal to that of one of the incoming neutrinos. Let us define $\Psi_i$ and $\Phi_i$ as the flavour wavefunction. There are two possible combinations,

$$\Psi_i^* \delta_{ij} \Psi_j \Phi_k^* \delta_{kl} \Phi_l = 1 \tag{3.2}$$

$$\Psi_i^* \delta_{il} \Phi_l \Phi_k^* \delta_{kj} \Psi_j = \frac{1}{2}(1 + \Psi_i^* \vec{\sigma}_{ij} \Psi_j \cdot \Phi_k^* \vec{\sigma}_{kl} \Phi_l). \tag{3.3}$$

where the following property was used,

$$2\delta_{il}\delta_{jk} = \delta_{ij}\delta_{kl} + \vec{\sigma}_{ij} \cdot \vec{\sigma}_{kl} \tag{3.4}$$

in Eq.(3.3).

Hence the complete flavour space Hamiltonian for two interacting neutrinos is proportional to,

$$\frac{3}{2} + \frac{1}{2}\,\vec{\sigma}_1 \cdot \vec{\sigma}_2. \tag{3.5}$$

Now consider the total spin operator for particles i and j,

$$\vec{S}_{ij} = \vec{s}_i + \vec{s}_j, \quad \text{where} \quad \vec{s} = \frac{1}{2}\vec{\sigma} \quad \text{and} \quad \vec{\sigma} \quad \text{are the Pauli matrices.} \tag{3.6}$$

From the theory of addition of angular momenta the interaction between pairs of spin is,

$$\begin{aligned} H_{ij} &= g\hat{S}_{ij}^2 = g(\vec{s}_i + \vec{s}_j)^2 \\ &= \frac{g}{2}\,(3 + \vec{\sigma}_i \cdot \vec{\sigma}_j) \end{aligned} \tag{3.7}$$

where g is the coupling. Comparing this Hamiltonian, describing two interacting spins, to the interaction of two neutrinos, Eq.(3.5), the equivalence of the spin system and neutrino system is clear. Henceforth we follow F&L and use a system of interacting spins to obtain information about the neutrino system.

Let the system initially contain N spin up particles and M spin down particles. The spin up particles can for instance represent electron neutrinos and the spin down particles can represent the muon neutrinos. Here, we construct the Hamiltonian in terms of interacting spins.

Using Eq.(3.7), the many-particle Hamiltonian is,

$$
\begin{aligned}
H_{int} &= g \sum_{i=1}^{M+N-1} \sum_{j=i+1}^{M+N} (\vec{s}_i + \vec{s}_j)^2 \\
&= g \left( \sum_{i=1}^{M+N-1} \sum_{j=i+1}^{M+N} 2\vec{s}_i \cdot \vec{s}_j + \frac{3}{4}(M+N)(M+N-1) \right).
\end{aligned}
\tag{3.8}
$$

We can compare our interaction Hamiltonian to the square of the spin operator for the whole system,

$$
\begin{aligned}
\hat{L} &= g \left( \sum_{i=1}^{N+M} \vec{s}_i \right)^2 \\
&= g \left( \sum_{i=1}^{M+N-1} \sum_{j=i+1}^{M+N} 2\vec{s}_i \cdot \vec{s}_j + \frac{3}{4}(M+N) \right)
\end{aligned}
\tag{3.9}
$$

which has eigenvalues (from angular momentum theory),

$$
E(L) = gL(L+1) \quad \text{with} \quad |N/2 - M/2| < L < N/2 + M/2 \quad .
\tag{3.10}
$$

Therefore our interaction Hamiltonian becomes,

$$
H_{int} = g[\hat{J}^2 + \frac{3}{4}(M+N)(M+N-2)]
\tag{3.11}
$$

with eigenvalues,

$$E(J, N, M) = g[J(J+1) + \frac{3}{4}(M+N)(M+N-2)] \quad \text{with} \quad |N/2 - M/2| < J < N/2 + M/2 \ .$$
$$(3.12)$$

Here $J$ is total angular momentum of the system.

In the neutrino system, the coupling depends upon the relative angle between the neutrino momenta, $\Theta$, namely $g = \frac{\sqrt{2}G_F}{V}(1 - \cos\Theta)$ where $G_F$ is the Fermi constant and $V$ is the volume of the box. However, for the system to be solved analytically the coupling must be constant. Hence we ignore the angular dependence and take the coupling to be,

$$g = \frac{\sqrt{2}G_F}{V}.$$
$$(3.13)$$

## 3.2 Constructing the many body density matrix

We choose to study the system of many neutrinos by constructing a density matrix. It is convenient to work in the total angular momentum (J) basis as the Hamiltonian is diagonal in this basis.

Our system begins in the state $|\, j_N, m_N \,\rangle \otimes |\, j_M, m_M \,\rangle$. Here, $j_N = N/2$ is the total angular momentum of all the spin up particles and $m_N = N/2$ is the projection along the $\hat{z}$ direction. Also, $j_M = M/2$ is the angular momentum of all the spin down particles and $m_M = -M/2$ is the projection. Rotating the initial state $|\, j_N, m_N \,\rangle \otimes |\, j_M, m_M \,\rangle$ to the total angular momentum (J) basis and evolving it to time t, we have,

$$| \, S(t) \, \rangle = | \, j_N, m_N \, \rangle \otimes | \, j_M, m_M \, \rangle (t) = \sum_{J=|j_N-j_M|}^{j_N+j_M} e^{-itE(J,N,M)}$$
$$\langle \, j_N, j_M, J, m \, | \, j_N, j_M, m_N, m_M \, \rangle$$
$$| \, j_N, j_M, J, m \, \rangle, \qquad (3.14)$$

where $\langle \, j_N, j_M, J, m \, | \, j_N, j_M, m_N, m_M \, \rangle$ is the Clebsch-Gordan coefficient where $j_N$ and $j_M$ are coupled to the total angular momentum of the system $J$ with projection in the $\hat{z}$ direction of $m = m_N + m_M$.

The density matrix is defined as,

$$\rho(t) = | \, S(t) \, \rangle \langle \, S(t) \, |. \qquad (3.15)$$

Hence we have,

$$\rho(t) = \sum_{J=|j_N-j_M|}^{j_N+j_M} \sum_{J'=|j_N-j_M|}^{j_N+j_M} e^{-it\Delta E(J,J')} \langle \, j_N, j_M, J, m \, | \, j_N, j_M, m_N, m_M \, \rangle$$
$$\langle \, j_N, j_M, J', m \, | \, j_N, j_M, m_N, m_M \, \rangle$$
$$| \, j_N, j_M, J, m \, \rangle \langle \, j_N, j_M, J', m \, |. \qquad (3.16)$$

where $\Delta E(J, J') = g[J(J+1) - J'(J'+1)]$ is the difference between the eigenvalues $E(J, N.M)$ and $E(J', N, M)$.

## 3.3   The Probability of the first spin being up

We are interested in finding the probability of one of the particles remaining in the spin up state if it was initially in the spin up state. This enables us to compare our results to [4] and also to analyse the flavour evolution of the system. The approach we use to find the probability is to "split off" the first spin from the system, so that the remaining $N + M - 1$ left over spins form a "background" with which this spin interacts. Therefore we first couple together the $N + M - 1$ remaining spins and then couple this to the first spin resulting in the total angular momentum, J. In a nut shell we begin by changing the way we couple the spins to the total angular momentum of the system so that it is then possible to "isolate" the first spin.

To achieve the recoupling it is convenient to again change the basis, this time to $|\frac{1}{2}, \lambda\rangle \otimes |j, \mu\rangle$.   $|\frac{1}{2}, \lambda\rangle$ is the state of one of the spin-1/2 particles with projection $\lambda$ and $|j, \mu\rangle$ is the state of M+N-1 remaining particles with angular momenta j and projection $\mu$. The density matrix in the new basis is constructed in the first subsection and then the probability is found in the second subsection.

### 3.3.1   Construction of the density matrix in the basis $|\frac{1}{2}, \lambda\rangle \otimes$ $|j, \mu\rangle$

We start with the state of Eq. (3.14), which is reproduced here for convenience, then transform this state to the new basis. We will omit the limits on the summation signs for the next few equations but we will comment on these later.

$$| \ j_N, m_N \ \rangle \otimes | \ j_M, m_M \ \rangle(t) = \sum_J e^{-itE(J,N,M)} \qquad \langle \ j_N, j_M, J, m \ | \ j_N, j_M, m_N, m_M \ \rangle$$
$$| \ j_N, j_M, J, m \ \rangle. \qquad (3.17)$$

To be able to change to the preferred basis $(|\frac{1}{2}, \lambda\rangle \otimes |j, \mu\rangle)$ we need to modify how the angular momenta couple to the total angular momenta, J. This can be done in anyway that is convenient. We choose to couple one of the spin up particles with angular momentum $\frac{1}{2}$ and the remaining $M + N - 1$ spin half particles with angular momentum $j$ to the total momenta of all the spin half particles, J. In order to do this we have specified that the momentum $j_N$ is a result of coupling one of the spin up particles with momenta $\frac{1}{2}$ and all the other $N - 1$ spin up particles with momenta $k = j_N - \frac{1}{2}$ (see figure 3.3.1 for a graphical representation). Hence $j$ is the result of coupling $j_M$ and $k$ (see figure 3.2) . Figure 3.3 represents the total recoupling.

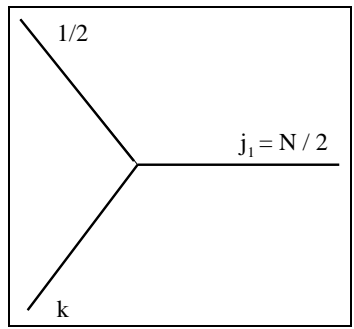

Figure 3.1: We couple one of the spin up particles with momenta $\frac{1}{2}$ and all the remaining spin up particles with momenta $k = \frac{N-1}{2}$, to the total momenta of all the spin up particles, $j_N = \frac{N}{2}$ .

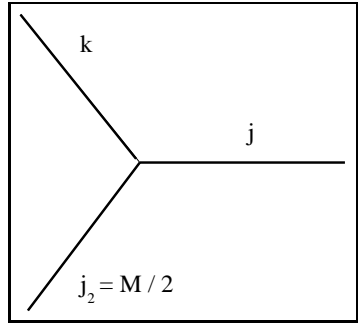

Figure 3.2: We couple k (the momenta of N-1 particles) and $j_M = M/2$ (the momenta of all the spin down particles (M) to j ).

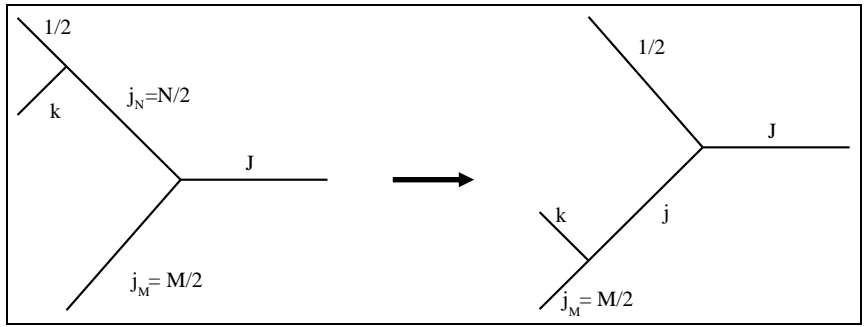

Figure 3.3: Graphical representation of recoupling. The diagram to the left of the arrow represents the original coupling: the momenta of all the spin up particles ($j_N$) and the momenta of all the spin down particles ($j_M$) were coupled to the total momenta of the system ($J$). The diagram to the right of the arrow represents the new coupling (see main text for description). Note that $k$ is the momentum of $N-1$ spin up particles.

The state $\mid j_N, j_M, J, m \rangle$ is represented in the new basis as,

$$\mid j_N, j_M, J, m \rangle = \sum_j \langle \frac{1}{2}, (k\, j_M)j; J \mid (\frac{1}{2}\, k)j_N, j_M; J \rangle \mid \frac{1}{2}, (k\, j_M)j; J, m \rangle. \quad (3.18)$$

The notation $(a\, b)c$ indicates that a and b couple to c. Here the recoupling coefficient,

$$\langle \frac{1}{2}, (k\, j_M)j; J \mid (\frac{1}{2}\, k)j_N, j_M; J \rangle \quad (3.19)$$

is proportional to the 6-j coefficient,

$$\langle \frac{1}{2}, (k\, j_M)j; J \mid (\frac{1}{2}\, k)j_N, j_M; J \rangle \equiv (-1)^{\frac{1}{2}+k+J+j_M} [(2j_N+1)(2j+1)]^{\frac{1}{2}} \left\{ \begin{array}{ccc} \frac{1}{2} & k & j_N \\ j_M & J & j \end{array} \right\}.$$

(3.20)

The left side of this coefficient represents the original coupling (the right side of figure 3.3). The right side represents the changed coupling (the left side of figure 3.3). The 6-j coefficient is a consequence of changing the coupling.

Rotating from the basis $\mid \frac{1}{2}, (k\, j_M)j; J, m \rangle$ to the basis $\mid \frac{1}{2}, \lambda \rangle \otimes \mid j, \mu \rangle$ and substituting $k = j_N - 1/2$, we have,

$$\mid j_N, j_M, J, m \rangle = \sum_{j,\lambda,\mu} (-1)^{J+j_M+j_N} [(2j_N+1)(2j+1)]^{\frac{1}{2}} \langle \frac{1}{2}, j, \lambda, \mu \mid \frac{1}{2}, j; J, m \rangle$$
$$\left\{ \begin{array}{ccc} \frac{1}{2} & j_N - \frac{1}{2} & j_N \\ j_M & J & j \end{array} \right\} \mid \frac{1}{2}, \lambda \rangle \otimes \mid j, \mu \rangle.$$

(3.21)

Finally substituting Eq. (3.21) into Eq. (3.17),

$$\mid j_N, m_N \rangle \otimes \mid j_M, m_M \rangle (t) = \sum_{J,j,\lambda,\mu} \exp^{-itE(J,N,M)} (-1)^{J+j_M+j_N} [(2j_N+1)(2j+1)]^{\frac{1}{2}}$$
$$\langle j_N, j_M, J, m \mid j_N, j_M, m_N, m_M \rangle \langle \frac{1}{2}, j, \lambda, \mu \mid \frac{1}{2}, j; J, m \rangle$$
$$\left\{ \begin{array}{ccc} \frac{1}{2} & j_N - \frac{1}{2} & j_N \\ j_M & J & j \end{array} \right\} \mid \frac{1}{2}, \lambda \rangle \otimes \mid j, \mu \rangle.$$

(3.22)

Note that the first particle is in the state up initially because we have specified that the total angular momenta of the spin up particles, $j_N$, is a result of coupling the angular momenta of one of the spin up particles and the angular momenta of the rest of the spin up particles, $k$.

It is now simple to construct the density matrix in this new basis using Eq.(3.22) and $\rho = \mid j_N, j_M, m_N, m_M \rangle \langle j_N, j_M, m_N, m_M \mid (t)$. We present the density matrix in component form,

$$
\rho_{\frac{1}{2}\, \lambda,\, j\, \mu\, ;\, \frac{1}{2}\, \lambda',\, j'\, \mu'}(t) = \sum_{J,J'} e^{-it\Delta E(J,J')} (-1)^{J-J'} (2j_N + 1)(2j + 1)
$$

$$
\langle j_N, j_M, J, m \mid j_N, j_M, m_N, m_M \rangle \langle j_N, j_M, J', m \mid j_N, j_M, m_N, m_M \rangle
$$

$$
\langle \frac{1}{2}, j, \lambda, \mu \mid \frac{1}{2}, j; J, m \rangle \langle \frac{1}{2}, j', \lambda', \mu' \mid \frac{1}{2}, j'; J', m \rangle
$$

$$
\left\{ \begin{array}{ccc} \frac{1}{2} & j_N - \frac{1}{2} & j_N \\ j_M & J & j \end{array} \right\} \left\{ \begin{array}{ccc} \frac{1}{2} & j_N - \frac{1}{2} & j_N \\ j_M & J' & j' \end{array} \right\}.
$$

$$
(3.23)
$$

## 3.3.2 Probability

The probability that the first spin remains in the up state can be found from the density matrix. In general the probability of the eigenvalue $a_i$ represented by the operator $A$ is $Tr(A\rho)$. In this case the operator is diagonal so that the probabilities of the z component of the spin are found on the diagonal of a density matrix. Hence the components of the density matrix which give probabilities are those where $j = j'$, $\mu = \mu'$, and $\lambda = \lambda'$ ie. $\rho_{\frac{1}{2}\, \lambda,\, j\, \mu; \frac{1}{2}\, \lambda,\, j\, \mu}$. Furthermore we are looking for the probability of the first spin remaining in the spin up state so that $\lambda = +\frac{1}{2}$. Using this information together with Eq. (3.23) we find the probability to be,

$$P_1(t) = \sum_{j,\mu} \rho_{\frac{1}{2}\,\frac{1}{2},\,j\,\mu;\frac{1}{2}\,\frac{1}{2},\,j\,\mu} = \sum_{J,J',j,\mu} e^{-it\Delta E(J,J')}(-1)^{J-J'}(2j_N+1)(2j+1)$$

$$\langle\, j_N, j_M, J, m \mid j_N, j_M, m_N, m_M \,\rangle$$

$$\langle\, j_N, j_M, J', m \mid j_N, j_M, m_N, m_M \,\rangle$$

$$\langle\, \frac{1}{2}, j, \frac{1}{2}, \mu \mid \frac{1}{2}, j; J, m \,\rangle \langle\, \frac{1}{2}, j, \frac{1}{2}, \mu' \mid \frac{1}{2}, j; J', m \,\rangle$$

$$\begin{Bmatrix} \frac{1}{2} & j_N - \frac{1}{2} & j_N \\ j_M & J & j \end{Bmatrix} \begin{Bmatrix} \frac{1}{2} & j_N - \frac{1}{2} & j_N \\ j_M & J' & j \end{Bmatrix}.$$

$$(3.24)$$

The third Clebsch-Gordan coefficient in Eq. (3.24), $\langle\, \frac{1}{2}, j, \frac{1}{2}, \mu \mid \frac{1}{2}, j; J, m \,\rangle$ represents the coupling of $\frac{1}{2}$ and $j$ to $J$. This gives $J - \frac{1}{2} \leqslant j \leqslant J + \frac{1}{2}$. In this same Clebsch-Gordan notice that we must have $\frac{1}{2} + \mu = m$, so that the summation over $\mu$ is unnecessary. Analysing the fourth Clebsch-Gordan coefficient, $\langle\, \frac{1}{2}, j, \frac{1}{2}, \mu' \mid \frac{1}{2}, j; J', m \,\rangle$, further simplifies the equation. This coefficient shows that we couple $\frac{1}{2}$ and $j$ to $J'$. Hence we have, $|j - \frac{1}{2}| \leqslant J' \leqslant j + \frac{1}{2}$. Now if $j = J + \frac{1}{2}$ then $J \leqslant J' \leqslant J + 1$, and if $j = J - \frac{1}{2}$ then $|J - 1| \leqslant J' \leqslant J$. Therefore $J' = |J - 1|, J, J + 1$.

Summing over $J'$,

$$P_1(t) = \sum_{J,j} (2j_N + 1)(2j + 1) \left| \langle\, j_N, j_M, J, m \mid j_N, j_M, m_N, m_M \,\rangle \right|^2$$

$$\times \left| \langle\, \frac{1}{2}, j, \frac{1}{2}, m - \frac{1}{2} \mid \frac{1}{2}, j; J, m \,\rangle \right|^2 \left| \left\{ \begin{array}{ccc} \frac{1}{2} & j_N - \frac{1}{2} & j_N \\ j_M & J & j \end{array} \right\} \right|^2$$

$$- (2j_N + 1)(2j + 1) e^{-it\Delta E(J+1,J)} \langle\, j_N, j_M, J, m \mid j_N, j_M, m_N, m_M \,\rangle$$

$$\times \langle\, j_N, j_M, J+1, m \mid j_N, j_M, m_N, m_M \,\rangle$$

$$\times \langle\, \frac{1}{2}, j, \frac{1}{2}, m - \frac{1}{2} \mid \frac{1}{2}, j; J, m \,\rangle \langle\, \frac{1}{2}, j, \frac{1}{2}, m - \frac{1}{2} \mid \frac{1}{2}, j; J+1, m \,\rangle$$

$$\times \left\{ \begin{array}{ccc} \frac{1}{2} & j_N - \frac{1}{2} & j_N \\ j_M & J & j \end{array} \right\} \left\{ \begin{array}{ccc} \frac{1}{2} & j_N - \frac{1}{2} & j_N \\ j_M & J+1 & j \end{array} \right\} \tag{3.25}$$

$$- (2j_N + 1)(2j + 1) e^{-it\Delta E(J,J-1)} \langle\, j_N, j_M, J, m \mid j_N, j_M, m_N, m_M \,\rangle$$

$$\times \langle\, j_N, j_M, |J-1|, m \mid j_N, j_M, m_N, m_M \,\rangle$$

$$\times \langle\, \frac{1}{2}, j, \frac{1}{2}, m - \frac{1}{2} \mid \frac{1}{2}, j; J, m \,\rangle \langle\, \frac{1}{2}, j, \frac{1}{2}, m - \frac{1}{2} \mid \frac{1}{2}, j; |J-1|, m \,\rangle$$

$$\times \left\{ \begin{array}{ccc} \frac{1}{2} & j_N - \frac{1}{2} & j_N \\ j_M & J & j \end{array} \right\} \left\{ \begin{array}{ccc} \frac{1}{2} & j_N - \frac{1}{2} & j_N \\ j_M & |J-1| & j \end{array} \right\}.$$

By shifting $J \rightarrow J+1$ in the third batch of expressions and noticing that particular

Clebsch-Gordan coefficients equal zero, namely,

$$\langle\, j_N, j_M, J+1, m \mid j_N, j_M, m_N, m_M \,\rangle \;=\; 0 \quad \text{when} \quad J = j_N + j_M \qquad (3.26)$$

$$\langle\, j_N, j_M, J, m \mid j_N, j_M, m_N, m_M \,\rangle \;=\; 0 \quad \text{when} \quad J = |j_N - j_M| - 1. \quad (3.27)$$

$$\langle\, \frac{1}{2}, j, \frac{1}{2}, m - \frac{1}{2} \mid \frac{1}{2}, j; J+1, m \,\rangle \;=\; 0 \quad \text{when} \quad j = J - \frac{1}{2}. \qquad (3.28)$$

$$\langle\, \frac{1}{2}, j, \frac{1}{2}, m - \frac{1}{2} \mid \frac{1}{2}, j; J-1, m \,\rangle \;=\; 0 \quad \text{when} \quad j = J + \frac{1}{2} \qquad (3.29)$$

we find,

$$
\begin{aligned}
P_1(t) \;=\; & \sum_{J=|j_N-j_M|}^{j_N+j_M} \sum_{j=J-\frac{1}{2}}^{J+\frac{1}{2}} \quad |\langle\, j_N, j_M, J, m \mid j_N, j_M, m_N, m_M \,\rangle|^2 \\
& \times \qquad\qquad \left| \langle\, \frac{1}{2}, j, \frac{1}{2}, m-\frac{1}{2} \mid \frac{1}{2}, j; J, m \,\rangle \right|^2 \qquad\qquad (3.30) \\
& \times \qquad\qquad \left| \left\{ \begin{array}{ccc} \frac{1}{2} & j_N - \frac{1}{2} & j_N \\ j_M & J & j \end{array} \right\} \right|^2 \\
& + \sum_{J=|j_N-j_M|}^{j_N+j_M-1} \qquad -(2j_N+1)(2j+1)\cos\left[t\,\mathbf{\Delta E(J+1, J)}\right] \\
& \times \qquad\qquad \langle\, j_N, j_M, J, m \mid j_N, j_M, m_N, m_M \,\rangle \\
& \times \qquad\qquad \langle\, j_N, j_M, J+1, m \mid j_N, j_M, m_N, m_M \,\rangle \qquad\qquad (3.31) \\
& \times \qquad\qquad \langle\, \frac{1}{2}, J+\frac{1}{2}, \frac{1}{2}, m-\frac{1}{2} \mid \frac{1}{2}, J+\frac{1}{2}; J, m \,\rangle \\
& \times \qquad\qquad \langle\, \frac{1}{2}, J+\frac{1}{2}, \frac{1}{2}, m-\frac{1}{2} \mid \frac{1}{2}, J+\frac{1}{2}; J+1, m \,\rangle \\
& \times \qquad\qquad \left\{ \begin{array}{ccc} \frac{1}{2} & j_N-\frac{1}{2} & j_N \\ j_M & J & J+\frac{1}{2} \end{array} \right\} \left\{ \begin{array}{ccc} \frac{1}{2} & j_N-\frac{1}{2} & j_N \\ j_M & J+1 & J+\frac{1}{2} \end{array} \right\}.
\end{aligned}
$$

Note that the probability constructed is completely general. That is, the probability is valid for any $j_N$, $m_N$, $j_M$ and $m_M$, with the restriction that the initial state of the first particle is the same as the initial state of the spin half particles that make up $j_N$.

Recall that the case we are interested in consists of the following quantum numbers,

$$j_N = \frac{N}{2} \quad m_N = \frac{N}{2} \tag{3.32}$$

$$j_M = \frac{M}{2} \quad m_M = -\frac{M}{2} \tag{3.33}$$

.

With these numbers one can further simplify the probability (Eq.3.30) by summing over $j$ evaluating the 6-j coefficients and the Clebsch-Gordan coefficients as well as substituting $\Delta E(J + 1, J)$. Equation (3.30) is then of the form,

$$P_1(t) = \sum_{J=|j_N-j_M|}^{j_N+j2} C(J) + \sum_{J=|j_N-j_M|}^{j_N+j2-1} \eta(J) cos[gt(2J + 2)] \tag{3.34}$$

where,

$$C(J) = \frac{(2J + 1)M\left[(M - N)^2(M + N + 2) - (4J(M - 3N)(J + 1))\right]\Gamma[M]\Gamma[N]}{16J(J + 1)\Gamma(\frac{M+N}{2} - J + 1))\Gamma(\frac{M+N}{2} + J + 2)}, \tag{3.35}$$

and

$$\eta(J) = \frac{M\left(J - \left(\frac{N-M}{2}\right) + 1\right)\left(J + \left(\frac{N-M}{2}\right) + 1\right)\Gamma(M)\Gamma(N)}{(J + 1)\Gamma\left(\frac{N+M}{2} - J\right)\Gamma\left(\frac{N+M}{2} + J + 2\right)}. \tag{3.36}$$

.

The constant in Eq.(3.35) is undefined for $J = 0$ which occurs only when $N = M$. Using Eq.(3.30) we find the constant for the case $J = 0$, is,

$$C(0) = \frac{1}{2(N + 1)}. \tag{3.37}$$

F&L who considered the special case $N = M$ found the same form for the proba-bility by using symmetries that are present only if $N = M$ to find $\sum_J C(J)$ and $\eta(J)$. For the case N=M our probability agrees exactly with the probability found by F&L.

# Chapter 4

# Analysis of a Simplified Many-Body Neutrino Model

This chapter consists of an analysis of the many-body neutrino model constructed in chapter 3. In particular we are interested in whether the one body approximation used in models for the early universe and supernova is valid. We investigate this question by analysing the probability that one of the spin up particles (or one of the neutrinos) immersed in a sea of many spins remains in the same spin up state (or in the same flavour eigenstate) at a later time. This allows us to track the flavour evolution of the neutrino. The signs that signal a breakdown of the one body approximation will be discussed in section 4.2.1 .

We begin the analysis by restating the probability for convenience and discussing some of the features found by studying plots of the probability for a variety of cases. These features are then thoroughly analysed.

## 4.1    The Features of the Probability

Recall that the probability of a particle initially in the spin up state (or a neutrino initially in a flavour eigenstate) remaining in the spin up state at a later time is,

$$P_1(t) = \sum_{J=|j_N-j_M|}^{j_N+j2} C(J) + \sum_{J=|j_N-j_M|}^{j_N+j2-1} \eta(J) cos[gt(2J+2)] \qquad (4.1)$$

with,

$$C(J) = \frac{(2J+1)M\left[\Delta^2 2\Sigma + 1\right) - (4J(M-3N)(J+1))\right]\Gamma[M]\Gamma[N]}{16J(J+1)\Gamma(\Sigma-J+1))\Gamma(\Sigma+J+2)} \qquad (4.2)$$

and,

$$\eta(J) = \frac{M\left(J-\Delta+1\right)\left(J+\Delta+1\right)\Gamma(M)\Gamma(N)}{(J+1)\Gamma\left(\Sigma-J\right)\Gamma\left(\Sigma+J+2\right)}, \qquad (4.3)$$

where we have defined,

$$\Delta = \frac{N-M}{2} \quad \Sigma = \frac{N+M}{2}. \qquad (4.4)$$

Note that Eq.(4.2) is undefined for $J=0$ when $N=M$. We use Eq.(3.30) to find,

$$C(0) = \frac{1}{2(N+1)}. \qquad (4.5)$$

Notice that the probability depends on the number of spin up and spin down particles due to the sum over J. We plot the probability according to Eq.(4.1) using various numbers of spin up particles, N, and spin down particles, M. The plots are shown in figures 4.1 and 4.2. In these figures the time is plotted in terms of the scaled time, $\tau = gt(N+M)$, so that we may compare our results to previous work in this area.

The main features of the solution are discussed below.

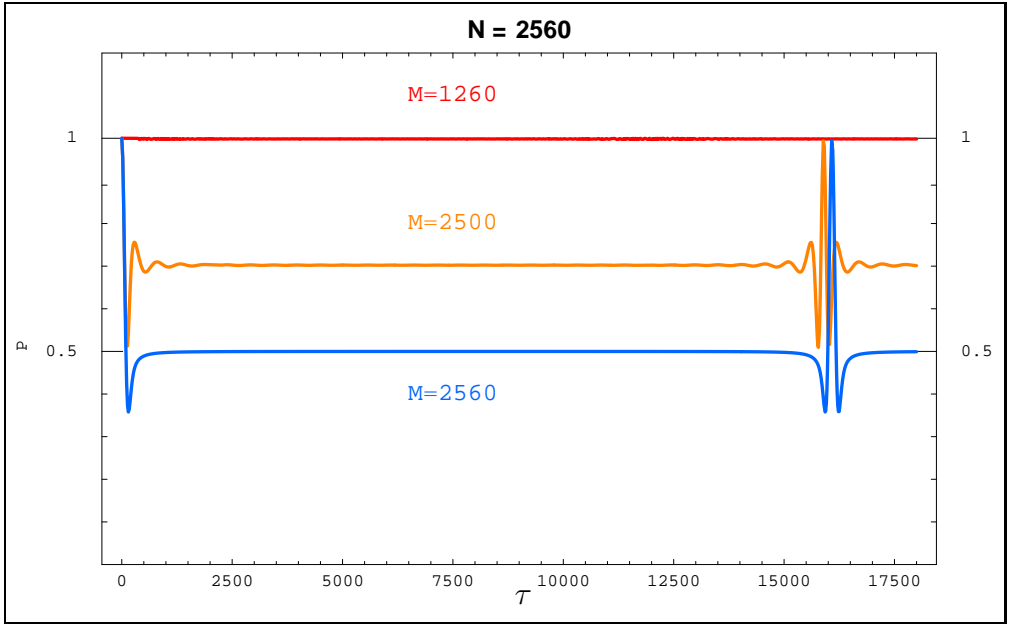

Figure 4.1: The probability of spin preservation for N=2560 spin up particles and various numbers of spin down particles, M. The time is scaled such that $\tau = gt(N+M)$

- The bottom two plots of figures 4.1 and 4.2 show that for $N \sim M$ the system equilibrates to the value of the average probability, $\bar{P}_1$, after some time. During the equilibration phase the system does not evolve until the wavetrain, seen for small t in the figures, reemerges much later. Indeed, F&L found the time for the system with $N = M$ to equilibrate is proportional to $g^{-1}N^{-\frac{1}{2}}$. We will discuss the significance of this in section 4.2.

- As $|N - M|$ increases the average probability, $\bar{P}_1$ increases. This is the case even if $N < M$. In fact, the lowest value is $\bar{P}_1 = \frac{1}{2}$, which occurs when $N = M$. This seems counter-intuitive at first. Indeed, it means that if we have a single spin, initially in the spin up state, coupled to a large sea of spin down particles, the single spin does not equilibrate to a state mostly oriented down, as one may expect, but remains aligned up.

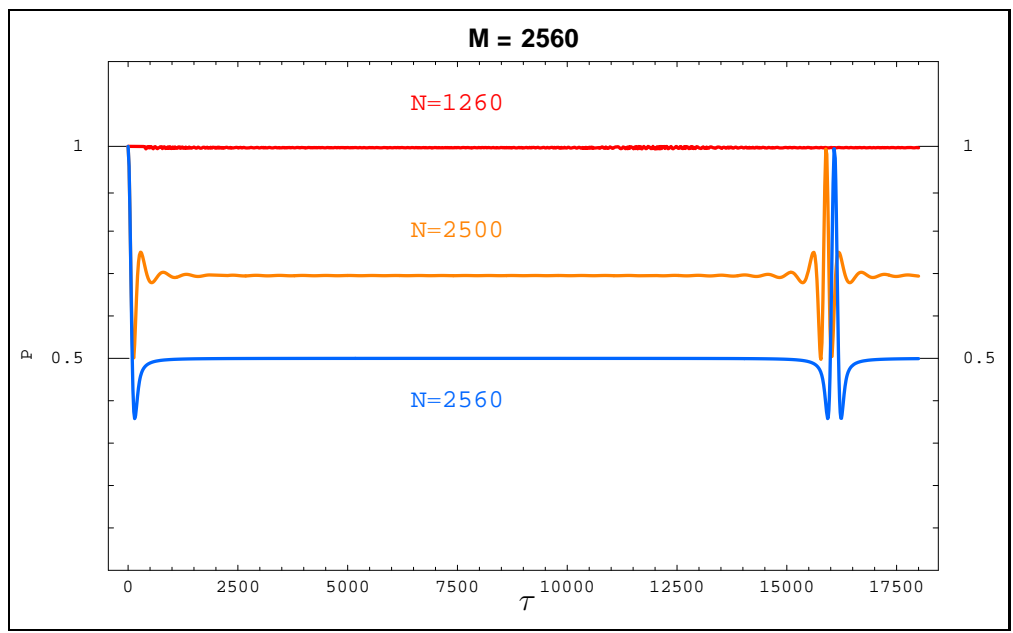

Figure 4.2: The probability of spin preservation for M=2560 spin down particles and various numbers of spin up particles, N. Note that even when $M > N$ the probability is above $\frac{1}{2}$. The time is scaled such that $\tau = gt(N + M)$.

- As $|N - M|$ increases the system begins to exhibit oscillation. The frequency increases with $|N - M|$ but the amplitude decreases as $|N - M|$ increases. As $|N - M|$ increases the system begin to show very little evolution due to the small amplitude of the oscillations. We call this the "freeze out" state. Freeze out is observed in the top plot of figures 4.1 and 4.2 . It is reproduced at higher resolution in figure 4.3 which shows the small amplitude and fast oscillations.

- The probability is periodic.

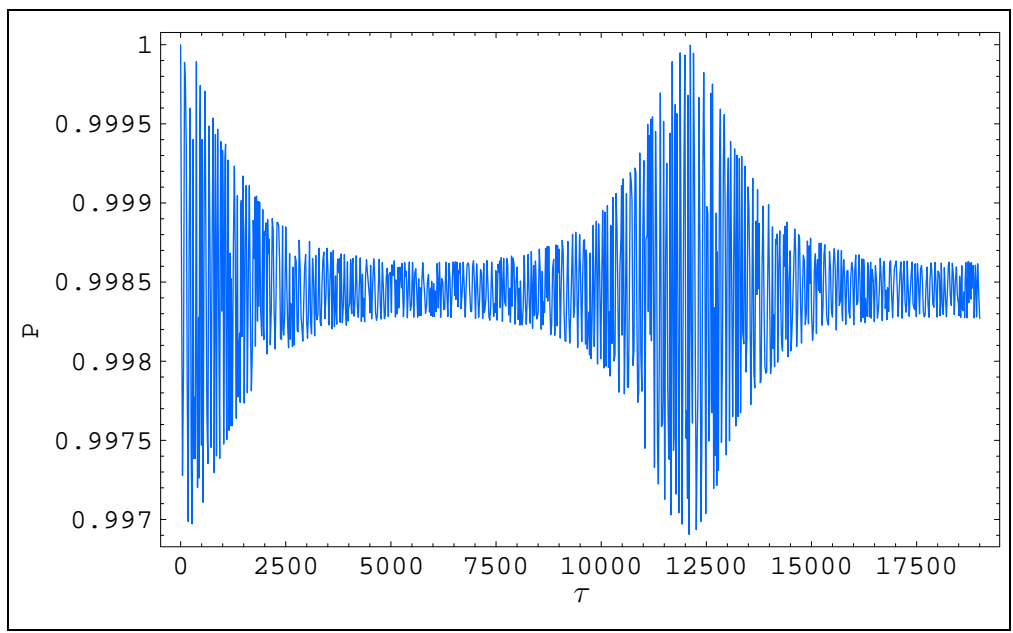

Figure 4.3: A close up of the plot $N = 2560$ and $M = 1280$. Note the vertical scale. The probability is very close to one and fluctuates minimally about the average value. The time is scaled such that $\tau = gt(N + M)$.

## 4.2 Equilibration

The equilibration time of the system is a tool which we can use to find if there is a breakdown in the one-body approximation. In this case, in which the neutrinos begin in flavour eigenstates, the one-particle description predicts time scales to be of an incoherent nature. Hence if one is able to find faster time scales such as that of a coherent nature this would indicate that the one-particle approximation is not valid in dense neutrino systems. To understand the analysis of the system we have constructed we must define the signs for coherent and incoherent time evolution.

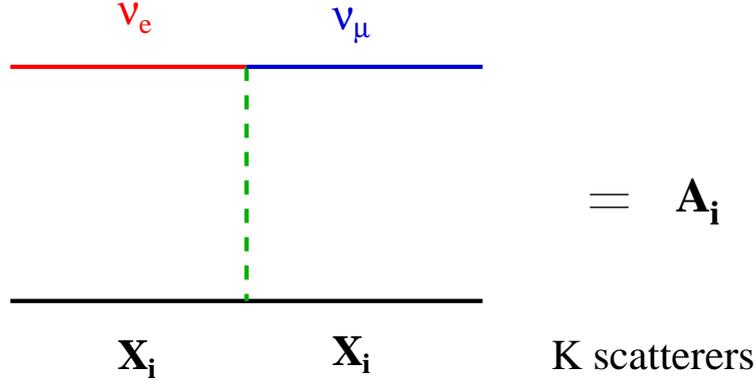

Figure 4.4: A Feynman diagram for neutrino flavour conversion due to the scattering of the beam neutrino, $\nu_e$, with the ith neutrino in the background, $X_i$. The amplitude for this event to occur is $A_i$. There are K neutrinos in the background

### 4.2.1 Signals for coherent and incoherent time evolution

Consider the Feynman diagram for neutrino flavour change in figure 4.4. This diagram depicts a neutrino in the beam interacting with the ith neutrino in the background. The amplitude for this event to occur is $A_i$. The probability of flavour conversion clearly depends on the number of neutrinos in the background, which we define as K, as well as the amplitude $A_i$. The neutrino in the beam can scatter off the background neutrinos either coherently or incoherently which lead to different probabilities for flavour conversion.

The probability of flavour change due to coherent scattering is,

$$P(\nu_e \to \nu_\mu) = \left| \sum_i A_i \right|^2 = K^2 |A|^2. \tag{4.6}$$

The probability of flavour change due to incoherent scattering is,

$$P(\nu_e \to \nu_\mu) = \sum_i |A_i|^2 = K|A|^2. \tag{4.7}$$

Cleary $A \propto t$, therefore the time scale on which the probability evolves is,

$$t_{coh} \quad \propto \quad (K)^{-1}, \tag{4.8}$$

$$t_{inc} \quad \propto \quad (K)^{-\frac{1}{2}} \tag{4.9}$$

for coherent scattering and incoherent scattering respectively. Therefore flavour conversion due to coherent scattering occurs much quicker than for incoherent scattering if the system contains a large number of particles. This is certainly the case for a real neutrino system.

The time scales in Eqs.(4.8) and (4.9) are thus the sign we must look for when analysing our probability. If we find that the equilibration time is inversely proportional to the number of particles then flavour conversion takes place faster than what is predicted by the one-body description. On the other hand, if we find that the equilibration time is inversely proportional to the square root of the number of particles then our analysis will agree with the one-body description.

### 4.2.2 Calculation of the equilibration time

Recall that equilibration can occur only if $|N - M|$ is small. The coefficient $\eta(J)$ in Eq.(4.1) can be usefully approximated for the case when $M - N$ is small. This approximation was found as a result of collaboration with Alexander Friedland.

We use the Stirling approximation, $\Gamma[z] = \sqrt{2\pi}e^{(z-\frac{1}{2})\ln(z-1)-(z-1)}$ and $\ln(x(1+\frac{a}{x})) \approx \ln x + \frac{a}{x}\left(1 - \frac{a}{2x}\right)$, to find that Eq.(4.3) reduces to,

$$\eta(J) \approx \frac{(J+1)^2 - \Delta^2}{(J+1)} \, e^{B - \frac{2(M+N+1)}{(M+N)^2}(J+1)^2} \tag{4.10}$$

where,

$$B = \left(M + \frac{1}{2}\right)\ln M + \left(N - \frac{1}{2}\right)\ln(N-1) - (M+N+1)\ln\Sigma + 1. \qquad (4.11)$$

For definiteness we take $M > N$ and approximate $M - N \ll N$, then further simplifying the natural logs by a series,

$$B \approx \left(N - \frac{1}{2}\right)\ln(N-1) - \left(N + \frac{1}{2}\right)\ln N + 1 + \frac{2N-1}{8N^2}(N-M)^2. \qquad (4.12)$$

Keeping only terms of order $\frac{1}{N}$ we find,

$$\eta(J) \approx \frac{(J+1)^2 - \Delta^2}{N(J+1)}\, e^{\frac{\Delta^2}{N}} e^{\frac{(J+1)^2}{\Sigma}}, \qquad (4.13)$$

where, as before $\Delta = \frac{N-M}{2}$ and $\Sigma = \frac{N+M}{2}$.

The above equation shows that $\eta(J)$ is a Gaussian centred about $J = -1$. The width of the Gaussian, $\sigma$ is related to the equilibration time by the uncertainty principle. The coefficient, $\eta(J)$, is the probability amplitude for a given angular momentum J. Since $\Delta E \approx g\sigma$ we can use the time-energy uncertainty principle to find the equilibration time. Note that in units of $\hbar = 1$ the time-energy uncertainty principle can be written as [44],

$$\left(\frac{\Delta\hat{A}}{\left(\frac{d\langle\hat{A}\rangle}{dt}\right)}\right)\Delta E \approx 1 \qquad (4.14)$$

Here $\hat{A}$ is some quantum mechanical operator. If we identify $\Delta t = \Delta A / \left(\frac{d\langle A\rangle}{dt}\right)$ then the meaning of $\Delta t$ is transparent. $\Delta t$ is the characteristic evolution time of the statistical distribution of A. Hence $1/g\sigma$ gives the equilibration time.

$$t_{eq} \propto g^{-1}\sqrt{2}(M+N)^{-1/2}. \tag{4.15}$$

The equilibration time is inversely proportional to the square root of the number of particles in the system hence it is of an incoherent nature.

To verify this argument we find an integral representation of $\sum_J \eta(J) \cos(2gt(J+1))$. To do this, we have used Eq.(4.13) together with $\cos(2gt(J+1)) = Re\left(e^{i2gt(J+1)}\right)$ and shifted the variable, J to $\tilde{J} = J - \Delta$.

$$\sum_J \eta(J) \cos(2gt(J+1) \approx \frac{e^{\frac{\Delta^2}{\Sigma+\Delta}}}{\Sigma+\Delta} \, e^{-\Sigma g^2 t^2} Re\left[\int_0^\infty \frac{(\tilde{J}+2\Delta)\tilde{J}}{(\tilde{J}+\Delta)} \, e^{-\left(\frac{\tilde{J}+\Delta}{\sqrt{\Sigma}}+\sqrt{\Sigma}igt\right)^2}\right]. \tag{4.16}$$

Since $|N-M|$ is small, we can Taylor expand the coefficient in the integral around $\Delta = 0$,

$$\frac{(\tilde{J}+2\Delta)\tilde{J}}{(\tilde{J}+\Delta)} = \tilde{J} + \Delta + \frac{\Delta^2}{\tilde{J}} + .... \tag{4.17}$$

Keeping only the first order term is a good approximation. The second order term, $\frac{\Delta^2}{J}$, only becomes important when $J > \Delta^2$ for sufficiently large numbers of particles. Taking the maximum value of the net angular momentum,$J = \frac{N+M}{2}$, gives the most stringent limit in the inequality. With this expression, the inequality becomes,

$$\sqrt{2(N+M)} > |N-M|. \tag{4.18}$$

Hence the second order term only becomes important when $|N-M|$ is large. In contrast, equilibration occurs when $|N-M|$ is small.

Substituting the Taylor series for the coefficient to first order into Eq.(4.16), defining $\tilde{J} - \Delta \equiv z$ and setting $|N - M| \approx 0$ in the lower limit of the integrand,

$$
\begin{aligned}
\sum_J \eta(J) \cos\left(2gt(J+1)\right) &\approx \frac{e^{\frac{\Delta^2}{\Sigma+\Delta}}}{\Sigma+\Delta} \, e^{-\Sigma g^2 t^2} \, Re\left[\int_0^\infty dz \, z \, e^{-\left(\frac{z}{\sqrt{\Sigma}}+\sqrt{\Sigma}igt\right)^2}\right] \\
&\approx \frac{\Sigma}{\Sigma+\Delta} \, e^{\frac{\Delta^2}{\Sigma+\Delta}} \int_0^\infty dz \, z \, e^{-z^2} \cos[2\sqrt{\Sigma}gtz]. \quad (4.19)
\end{aligned}
$$

The integral can now be calculated analytically. We use the contour in figure 4.5 to find the answer.

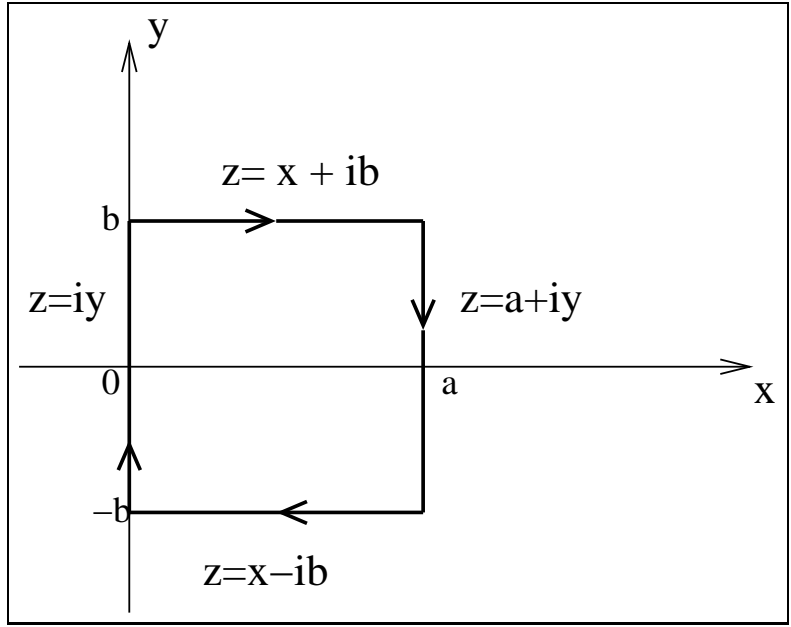

Figure 4.5: The contour used to solve the integral in Eq.(4.19).

By Couchy's Theorem,

$$
\int_a^0 e^{-(x-ib)^2}dx + \int_{-b}^b e^{-(iy)^2}idy + \int_b^{-b} e^{-(a+iy)^2}idy + \int_0^a e^{-(x+ib)^2}dx = 0. \quad (4.20)
$$

By letting $a \to \infty$ and rearranging, Eq.(4.20) becomes,

$$
\begin{aligned}
\int_0^\infty e^{-x^2} \sin(2bx)dx &= \frac{e^{-b^2}}{2} \int_{-b}^b e^{y^2} dy \\
&= \frac{e^{-b^2}}{2} \sqrt{\pi} \text{Erfi[b]},
\end{aligned}
\tag{4.21}
$$

where Erfi[b] is the imaginary error function,

$$
\text{Erfi[b]} = \frac{2}{\pi} \int_0^b e^{-y^2} dy.
\tag{4.22}
$$

Now taking the derivative of both sides Eq.(4.21) with respect to b we find,

$$
\int_0^\infty 2x \, e^{-x^2} cos(2bx) = \frac{1}{2}(1 - b\sqrt{\pi}e^{-b^2}\text{Erfi[b]}).
\tag{4.23}
$$

Identifying that $b = \sqrt{\Sigma}gt$, solves Eq.(4.19),

$$
\sum_J \eta(J) \cos[2gt(J+1)] \approx \frac{\Sigma}{2(\Sigma+\Delta)} e^{\frac{\Delta^2}{\Sigma+\Delta}} \left[ 1 - \sqrt{\pi\Sigma}gte^{-\Sigma g^2 t^2}\text{Erfi}[\sqrt{\Sigma}gt] \right].
\tag{4.24}
$$

Recall that $\Sigma = \frac{N+M}{2}$. Hence, the integral approximation shows the flavour equilibration time is,

$$
t_{eq} \propto (N + M)^{-1/2},
\tag{4.25}
$$

thus confirming Eq.(4.15).

Note also that in the case $N = M$, Eq.(4.24) reduces to,

$$
\sum_J \eta(J) \cos[2gt(J+1)] \approx \frac{1}{2} \left[ 1 - \sqrt{\pi N}gte^{-Ng^2 t^2}\text{Erfi}[\sqrt{N}gt] \right],
\tag{4.26}
$$

which is the integral approximation that F&L [4] found in their analysis.

## 4.3  Freeze-out

The average probability, $\bar{P}$, is the point that the probability oscillates about. This point is just the constant $\sum_J C(J)$. To investigate the behaviour of the average probability when we change the numbers of spin up and spin down, we plot $\bar{P} = \sum_J C(J)$ against $N - M$ while keeping $M + N$ constant. A plot for various constants is shown in figure 4.6.

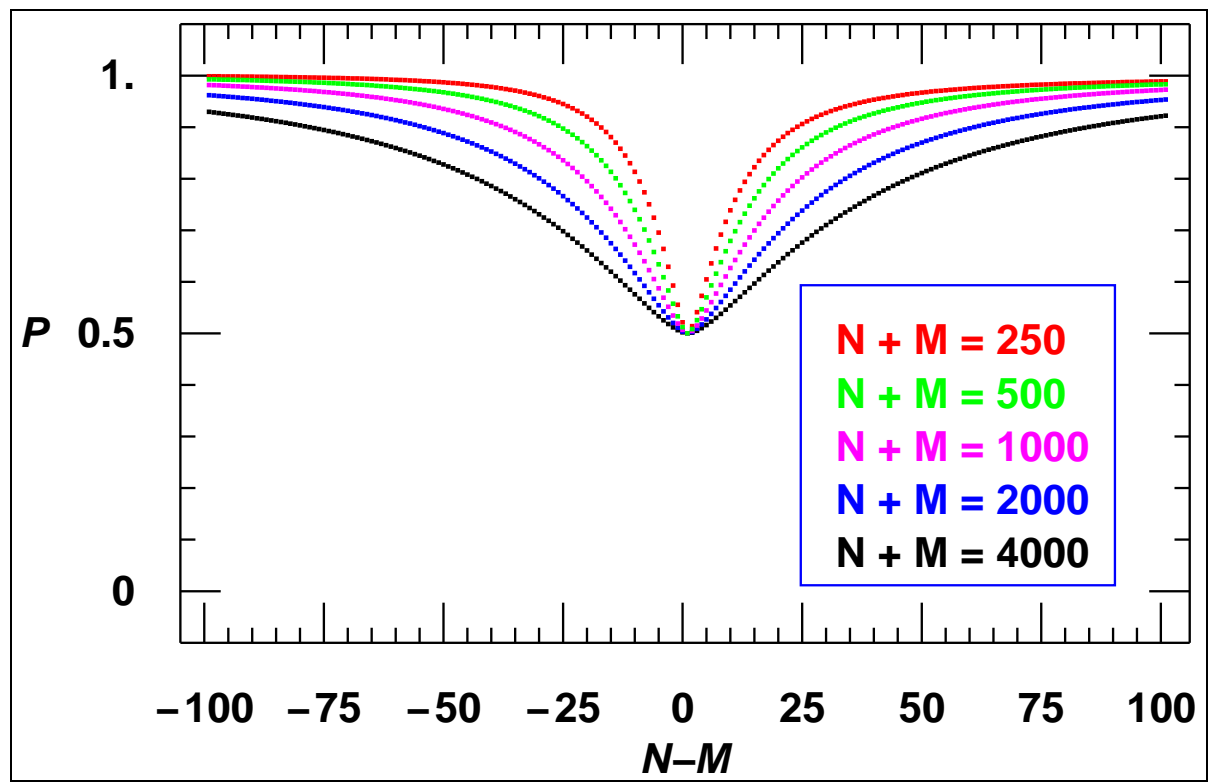

Figure 4.6: The average probability $(\bar{P}_1 = \frac{1}{2} + \sum_J C(J))$ when we change the number of spin ups and spin downs while keeping $M + N$ constant.

Figure 4.6 shows that as $|N - M|$ increases the average probability is close to 1. the region where the probability is approximately 1 is freeze-out i.e. there is little evolution in the system, for all other cases, namely when $N \sim M$, the system equilibrates. Note that the lowest value for the average probability is $\frac{1}{2}$ which is attained when $N = M$.

The puzzling feature in the freeze-out regime is that even if $N < M$ the probability of the spin remaining in the state up is approximately one. In this system all spins are interacting and the interaction causes an exchange of spin components hence we would expect the probability for the spin remaining in the spin up state, for the case $N < M$, to be close to 0. Further, freeze-out seems to occur for many more initial configurations of N and M than does equilibration. We investigate the reason for this behaviour below.

## 4.3.1   Another formula for the probability

The probability of one of the initial spin up particles remaining in the spin up state can be rewritten in a form that is more transparent for the interpretation of the freeze-out. Let us denote the state of the whole system by $|S(t)\rangle$. Then,

$$P_1(t) = \langle S(t)| + 1/2 \rangle \langle +1/2 | S(t) \rangle. \tag{4.27}$$

Now, let $s_z$ be the z component of the spin operator we are interested in. From quantum mechanics,

$$s_z = 1/2(| + 1/2 \rangle \langle +1/2 | - | - 1/2 \rangle \langle -1/2 |). \tag{4.28}$$

Using the completeness relation $| + 1/2 \rangle \langle +1/2 | + | - 1/2 \rangle \langle -1/2 | = 1$, Eq.(4.28) and

Eq.(4.27), yields,

$$P_1(t) = \frac{1}{2} + \langle S(t)|s_z|S(t)\rangle. \tag{4.29}$$

Observing that the expectation value $\langle S(t)|s_z|S(t)\rangle$ is equal to $\frac{1}{N}\langle S(t)| \ j_{Nz} \ |S(t)\rangle$, where $j_{Nz}$ is the z component of the angular momentum of all the spin up particles, $j_N$, the probability is

$$P_1(t) = \frac{1}{2} + \frac{1}{N}\langle S(t)| \ j_{Nz} \ |S(t)\rangle. \tag{4.30}$$

## 4.3.2  Interpretation of Freeze-out

**A qualitative analysis**

We can understand the freeze out by turning to the reduction of the system to two angular momenta and using a semi-classical treatment. Semi-classical analysis of angular momentum that this analysis is based on can be found in [45–47] .

All spins that start out in the same state must evolve in the same way. Hence we can combine all the spin up particles into one object, the angular momentum $j_N$, and all the spin down particles into one object, $j_M$, at any moment in time. The problem thus reduces to a system consisting of two coupled angular momenta.

Classically, the Hamiltonian for two interacting angular momenta is,

$$H_{cl} = kj_N \cdot j_M \tag{4.31}$$

with k being the coupling constant.

In classical mechanics the time evolution of the lth component of $j_N$ can be described via the following equation,

$$\begin{aligned}
\frac{dj_{Nl}}{dt} &= \{j_{Nl}, H\} \\
&= k(j_{Nl}, j_{Nm}) \cdot j_{Mm} \\
&= k\epsilon_{lmn} j_{Nn} j_{Mm}.
\end{aligned} \tag{4.32}$$

Here $\{\}$ is a Poisson bracket. Continuing with the derivation we find,

$$\frac{dj_N}{dt} = -k(j_N \times j_M). \tag{4.33}$$

Similarly,

$$\frac{dj_M}{dt} = k(j_N \times j_M). \tag{4.34}$$

Now,

$$\frac{d}{dt}(j_N \times j_M) = \frac{dj_N}{dt} \times j_M + j_N \times \frac{dj_M}{dt}. \tag{4.35}$$

Using equations (4.33) and (4.34) we find,

$$\begin{aligned}
\frac{d}{dt}(j_N \times j_M) &= k(j_N + j_M) \times (j_N \times j_M) \\
&= kJ_{cl} \times (j_N \times j_M).
\end{aligned} \tag{4.36}$$

Hence the two angular momenta, $j_N$ and $j_M$, precess about the classical total angular momentum vector, $J_{cl}$. We can also represent the evolution of the total angular momentum, $J$, by the precession of $J$ about the z-axis symbolising the quantumechanically fluctuating x and y components. The whole system is depicted in figure 4.7.

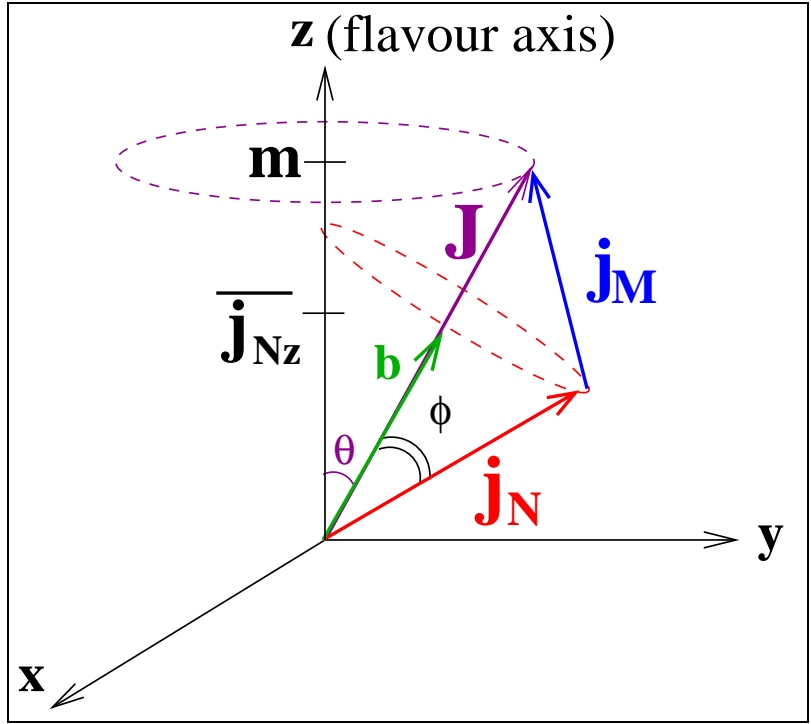

Figure 4.7: The total angular momenta vector, J, precesses about the z-axis. The angular momenta of all the spin up particles ($j_N$) and the total angular momenta of all the spin down particles ($j_M$) precess about the total angular momenta, J.

The system does not evolve significantly as long as $|N - M|$ is large. Therefore N is small relative to M and vice-versa. In this case the total range of total angular momenta J available in the system is small as $\frac{|N-M|}{2} < J < \frac{N+M}{2}$. Recall that the projection of J onto the z-axis is $m = \frac{N-M}{2}$, hence in this regime $m \approx |J|$. This means that that the x and y components of the total angular momenta, $J_{xy}$, are small or equivalently the angle $\theta$ in figure 4.7 is small. This can be proven mathematically. Since $J_z = m = |J|$ and $|J|^2 = J(J+1)$ then the equation for the circle described by J in the x-y plane is,

$$J^2 = J_x^2 + J_y^2 + J_z^2$$
$$J_x^2 + J_y^2 = J^2 - J_z^2 = J$$

$$(4.37)$$

Hence the radius of the circle, $J_{xy}$, is $\sqrt{J}$. Further to first order the length of the angular momentum vector is $J$. Therefore $|J| \gg J_{xy}$ as required. The circle described by J is small and therefore we can view the angular momentum vector as being almost stationary rather than precessing. Thus J behaves like a classical angular momentum vector in this regime.

Recall that the probability can be written as,

$$P = \frac{1}{2} + \frac{1}{N} \langle\, S(t) \mid j_{Nz} \mid S(t)\, \rangle \qquad (4.38)$$

The expression $\langle\, S(t) \mid j_{Nz} \mid S(t)\, \rangle$ is the expectation value of the z component of $j_N$. In figure 4.7, $\langle\, S(t) \mid j_{Nz} \mid S(t)\, \rangle$ is $\overline{j_{Nz}}$. Classically, we find $\overline{j_{Nz}}$ by projecting $j_N$ onto $J$, which gives us the vector b in figure 4.7, then projecting b onto the z-axis. Since $\theta$ is small then we expect $\langle\, S(t) \mid j_{Nz} \mid S(t)\, \rangle \approx \frac{N}{2}$ and thus $P \approx 1$ according to Eq.(4.38) (which is now a semi-classical expression as we have used a classical derivation of $\langle\, S(t) \mid j_z^1 \mid S(t)\, \rangle$).

This analysis becomes invalid once the total angular momentum vector, J, is no longer "pinned" to the z axis. This occurs when the length of J is larger than its projection onto the z axis, m, or equivalently when the x and y components of J become larger than $J_z = m$,

$$|J| > m \tag{4.39}$$

$$\sqrt{\frac{N+M}{2}} > \frac{|N-M|}{2} \tag{4.40}$$

This is thus the condition for which equilibration occurs. Notice that the width of the equilibration region of figure 4.6 is indeed $\sqrt{2(N+M)}$.

## 4.4   The Period of the Probability

In this section we will use the scaled time $\tau$ and alert the reader if we are using unscaled time, t. This section contains a detailed derivation of the period. The final result is stated in Eq.(4.56).

In the scaled time $\tau$, the probability is,

$$P(\tau) = \sum_{J=|j_N - j_M|}^{j_N + j2} C(J) + \sum_{J=|j_N - j_M|}^{j_N + j2 - 1} \eta(J) \cos\left[\tau \frac{2J+2}{N+M}\right] \tag{4.41}$$

The periodicity is a consequence of the probability being a finite sum of cosines. Each cosine in the sum satisfies,

$$\cos\left[\tau \frac{2J+2}{N+M}\right] = \cos\left[(\tau + T_J) \frac{2J+2}{N+M}\right] \tag{4.42}$$

Here $T_J$ is the period of the cosine corresponding to angular momenta J. Now for all J,

$$T_J = 2\pi \left(\frac{N+M}{2J+2}\right) \tag{4.43}$$

To find the period we need the least common multiple of the $T_J$'s.

**The Case for N=1 or M=1**

We first consider the case when N or M is one. Consider $N = 1$. For this case there is only one cosine term with $J = \frac{M}{2} - \frac{1}{2}$. Therefore Eq. (4.43) reduces to,

$$T_J = 2\pi \qquad (4.44)$$

If we take $M = 1$ instead, the result remains the same. Hence the period is $2\pi$ if $M = 1$ or $N = 1$.

**The Case for $N > 1$ and $M > 1$**

Here we consider the case when $N > 1$ and $M > 1$. The difficulty lies in the fact that we have to take into account the whole range of J. However this can be simplified by dissecting how the cos terms behave in relation to one another.

As we sum from $J = |j_N - j_M|$ to $J = j_N + j_M - 1$ the argument of each cosine increases and the period of each $\cos\left[\tau \frac{2J+2}{N+M}\right]$ decreases. Hence we only need to find when the periods of the cosine with the largest period, $\cos\left[\tau \frac{2(|j_N - j_M|)+2)}{N+M}\right]$, and the cosine with the smallest period, $\cos\left[\tau \frac{2(j_N + j_M - 1)+2}{N+M}\right]$ coincide.

In the following proof we let $N > M$. Note that if we take $M > N$ the proof is the same but M and N are swapped and the results remain same. Recall again $j_N = \frac{N}{2}$ and $j_M = \frac{M}{2}$.

The periods of the two cosine terms we are interested in are,

$$T_{J=\frac{N}{2}-\frac{M}{2}} = 2\pi \left( \frac{N+M}{N-M+2} \right) \qquad (4.45)$$

$$T_{J=\frac{N}{2}+\frac{M}{2}-1} = 2\pi \qquad (4.46)$$

Hence,

$$\cos\left[ \tau \, \frac{N-M+2}{N+M} \right] = \cos\left[ \left\{ \tau + \left( 2n'\pi \frac{N+M}{N-M+2} \right) \right\} \left( \frac{N-M+2}{N+M} \right) \right] \quad \text{where n}' = 0, 1, 2... \quad (4.47)$$

$$\cos\left[ \tau \right] = \cos\left[ \tau + 2n\pi \right] \quad \text{where n} = 0, 1, 2... \qquad (4.48)$$

Therefore the periods coincide when,

$$2n\pi = 2n'\pi \, \frac{N+M}{N-M+2} \quad \text{or} \qquad (4.49)$$

$$n = n' \, \frac{N+M}{N-M+2}. \qquad (4.50)$$

Both $n$ and $n'$ are restricted to being integers. Since we are interested in finding the period we only need to find the first time when both $n$ and $n'$ are integers simultaneously. This depends on N and M.

**CASE 1:** If $N+M$ is even (or J is an integer) then,

$$n' = \frac{N-M+2}{2}. \qquad (4.51)$$

Note that for any $N$ and $M$ if $N-M$ is even $n' \epsilon \mathbb{Z}$. Substituting this into Eq. (4.50),

$$n = \frac{N + M}{2} \tag{4.52}$$

**CASE 2:** If $N + M$ is odd (or J is half an integer) then,

$$n' = N - M + 2 \qquad \text{and} \tag{4.53}$$

$$n = (N + M) \tag{4.54}$$

Note that $N, M \epsilon \mathbb{Z}$ so that $n' \epsilon \mathbb{Z}$.

The recurrence time of $cos[\tau]$ (from Eq.(4.48)) is $2n\pi$. Hence substituting Eq. (4.52) and Eq. (4.54) into $2n\pi$ and changing back to unscaled time, we find the period to be,

$$T = \begin{cases} \frac{\pi}{g}, & \text{if } N + M = \text{even (or J=integer)} \\ \frac{2\pi}{g}, & \text{if } N + M = \text{odd (or J= half integer)} \\ \frac{2\pi}{g(N+M)}, & \text{if N=1 or M=1} \end{cases} \tag{4.55}$$

The discontinuity between the period when $N = 1$ or $M = 1$ and the other cases arises because in the first two cases there is an interference of many cosine waves (as there the cosines are summed over) and in the last case there is only one cosine wave. The period written in the form of Eq.(4.55) can be re-written to analyse the system in terms of the number density rather than the size of the box. Recall that $g = \frac{\sqrt{2} G_F}{V}$, then,

$$T = \begin{cases} \frac{(M+N)\pi}{\sqrt{2}\rho_\nu G_F}, & \text{if } N+M = \text{even (or J=integer)} \\\\ \frac{2(M+N)\pi}{\sqrt{2}\rho_\nu G_F}, & \text{if } N+M = \text{odd (or J= half integer)} \\\\ \frac{2\pi}{\sqrt{2}\rho_\nu G_F}, & \text{if N=1 or M=1} \end{cases} \qquad (4.56)$$

Here $\rho_\nu$ is the neutrino number density. In this form it is obvious that the period is proportional to the number of particles (except for the case $N = 1$ or $M = 1$) and inversely proportional to the number density. Hence the probability is non-trivially tied to the number of particles and the number density. In contrast the probability for flavour conversion in a system with $\nu - e^-$ scattering is dependent only on the density of the electrons. Furthermore the period (or quasi-period) of a many-particle system is generally related to the Poincare time which is proportional to $(N + M)!$. The period of this system is much smaller. Our results for the period agree with that of F&L for the case $M = N$.

As a final comment note that the approximation to the probability when $|N - M|$ is small, Eq.(4.26), destroys the periodicity as we take the upper limit of the integrand to infinity.

### 4.4.1    A note about minima

Note that for the case $N + M = $ even, the probability can never equal $\sum_J C(J) - \sum_J \eta(J)$, which is lowest possible value that the probability can be. We will call this a perfect minimum. For the perfect minimum to occur we must have, $\cos\left[\tau \frac{2J+2}{N+M}\right] = -1$ simultaneously for all J.

As previously stated, since the period of each cosine decreases as J increases, we only need to find the time when the cosine with the largest period and the cosine with the smallest period are simultaneously equal to $-1$. We set these two particular cosines equal to each other to find the times when they are equal.

$$\cos\left[\tau\frac{N-M+2}{N+M}\right] = \cos[2n\pi - \tau] \quad \text{where n} = 0, 1, 2... \qquad (4.57)$$

Hence the times when the two cosines are equal is,

$$\tau = \left(\frac{N+M}{N+1}\right)n\pi \qquad (4.58)$$

Substituting this time into the cosine with the smallest period, $cos[\tau]$, and setting it equal to $-1$ (as this is when the cosine is a minimum),

$$\cos[\frac{N+M}{N+1}\,n\pi] = -1 \quad \text{where} \quad n = 0, 1, 2... \qquad (4.59)$$

$$\left(\frac{N+M}{N+1}\right)n = 2n'+1 \quad \text{where} \quad n' = 0, 1, 2... \qquad (4.60)$$

Hence $\left(\frac{N+M}{N+1}\right)n$ must be an odd integer. For this term to be an integer the smallest value that $n$ can be is $N+1$. So we have,

$$n' = \frac{N+M-1}{2} \qquad (4.61)$$

Recall that $n'$ is also restricted to being an integer. Therefore $N+M-1$ must be even and so $N+M$ must be odd. Hence if $N+M =$ even there will never be a perfect minimum. If $N+M$ is odd it is straightforward to show that the times when the probability attains a perfect minimum is,

$$T_{min} = \frac{k\pi}{g} = \frac{k\pi(M+N)}{\sqrt{2}G_F\rho_\nu} \quad \text{where} \quad k = 1, 2, 3... \tag{4.62}$$

Note that this is half way between perfect maximums ( the case where $\cos[\tau\frac{(2J+2)}{(M+N)}] = 1$ simultaneously for all J and $P_1(t) = \sum_J C(J) + \sum_J \eta(J)$). Hence if $M + N$ is odd (or $J$ is half an integer) the probability has both a perfect minimum and a perfect maximum. This result shows yet another intriguing physical feature of the spin system. While a system with $N + M =$ even is characterized by a set of recurring maxima, in the system with $N + M =$ odd every other such maximum is replaced by a minimum. This behaviour is illustrated in figure 4.8. Note that in this figure the time is scaled so that $\tau = gt(M + N)$.

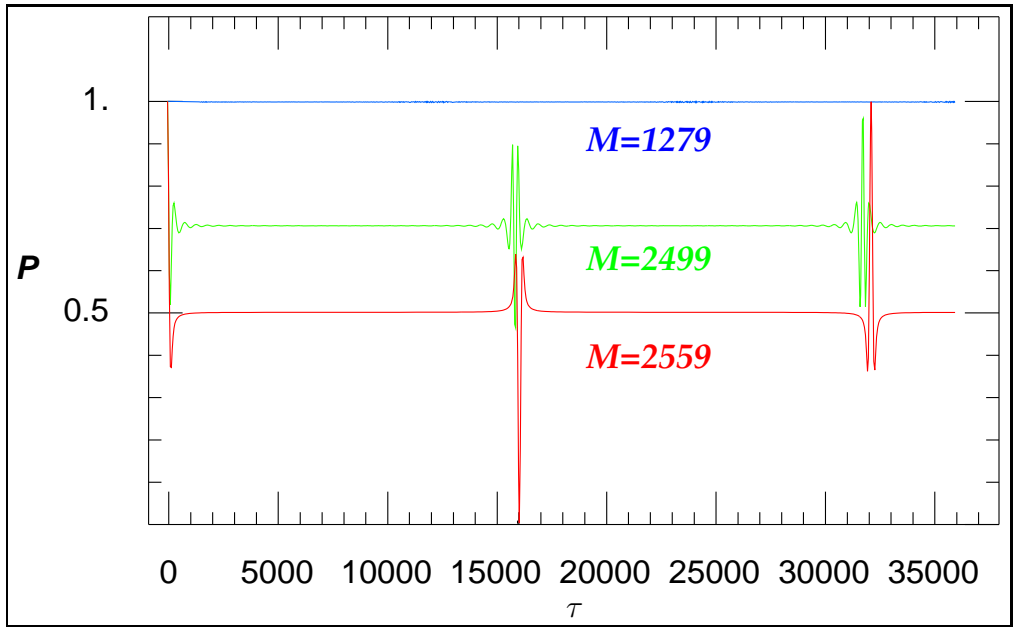

Figure 4.8: Plots of N=2560 and various numbers of spin down particles, M. The time is scaled so that $\tau = gt(M + N)$. Note that for all the graphs $N + M =$odd

## 4.5   Time Scales

As mentioned in section 4.1, as $|N - M|$ increases the system goes from equilibration at N=M to equilibration plus oscillation when $N \sim M$ to oscillations with a small amplitude when $|N - M| > \sqrt{2(N + M)}$. Further the system is periodic. Hence there are three time scales in this model which govern equilibration, oscillations and the period.

The one body description of the system employed by physicists studying dense neutrino systems predicts only one time scale due to incoherent scattering. Given that we have found three time scales in the system, does this mean that the one body description does not give enough information about the system and hence is not a valid approximation?

In the region $N \sim M$ (including $N = M$), we have found that equilibration occurs after $t = g(\sqrt{N + M})^{-\frac{1}{2}}$. This time is inversely proportional to the square root of the the number of particles hence it is of an incoherent nature as found in the one-body approximation. However as $|N - M|$ increases beyond 0 the system starts exhibiting oscillations. In section 4.2 we derived the equilibration time by using an approximation to the probability resulting in Eq.(4.26). We used a Taylor series around $|N - M| = 0$ and utilised only the first order terms. This approximation showed no oscillations in the system only the incoherent time scale. Presumably the oscillations would appear in the equation and thus become important when the approximation breaks down which we have found was when $|N - M| > \sqrt{2(N + M)}$. This is just the condition for freeze-out. Therefore the oscillations are only important in the freeze-out regime.

We have found that the solution to the system is periodic. For the case $N = M$ F&L, [4], showed that the period is longer than the lifetime of the universe: for a volume $V = 1$ cm$^3$ they found $T \sim 10^{22}$. In this system the period is constant (depending only on the coupling g) up to a factor of 2 controlled by whether $N + M$ is odd or even, hence for all cases it is the same as $N = M$. In the regime $N \sim M$ the period is also much smaller than the incoherent nature of the equilibration time. We can therefore conclude that the period is unlikely to play a major role in the neutrino system. In the context of the spin system the periodicity can be of importance. In fact, the periodicity is reminiscent of "spin echo" in actual spin systems.

In the freeze-out region the oscillations become important as discussed in the previous paragraph. Recall that $j_N$ precesses around the direction of total angular momentum, J. Since the projection of $j_N$ onto the z-axis controls the probability, this precession is the cause of the oscillations. The equation for the evolution of $j_N$, Eq.(4.36), shows that the frequency of the oscillations is $g|J_{cl}| = \frac{g}{2}|N - M|$. The frequency is linear in the number of particles and therefore it is faster than time scales due to incoherent scattering which are proportional to the square root of the number of particles. Can we reconcile this with the slower flavour evolution found in the one-particle model? The answer is yes. In the freeze-out regime the amplitude of the oscillations is small so that there is little evolution in the system. Due to the small amplitude the period is also not an important factor. In subsequent collaboration with Alexander Friedland, he found that the probability can be approximated very well in the region $\sqrt{N + M} < |N - M| \ll (N + M)$ by,

$$P_1(t) \;\simeq\; 1 - \frac{\Sigma}{2} \left( \frac{1}{\Delta^2} - \frac{\cos\left(2gt\Delta + 2\arctan\left(\frac{gt\Sigma}{\Delta}\right)\right)}{\Delta^2 + g^2 t^2 \Sigma^2} \right). \qquad (4.63)$$

This equation shows that the amplitude of the oscillations is $\frac{N + M}{|N - M|^2}$ showing that as $|N - M|$ increases the amplitude decreases and that the frequency of oscillations is $|N - M|$. Note that in the equilibration region $|N - M| \sim \sqrt{N + M}$, thus the oscillation time scale is indistinguishable from the incoherent equilibration time scale.

## 4.6   Summary

We have found that the neutrino system can be solved analytically by solving the equivalent spin system. We analysed the system using the probability of one of the initial spin up particles remaining in the spin up state. The probability has shown different behaviours in two regimes.

In the region $N \sim M$ the probability equilibrates to the average probability after a time, $t_{eq} \propto g^{-1}(N + M)^{-\frac{1}{2}}$ which is of an incoherent structure. Although there are oscillations in this regime (except for $N = M$) they have shown to be negligible. There is no evolution in the system until the wavetrain, seen before the equilibration, re-emerges due to periodicity of the probability. However this period is larger than the lifetime of the universe and the incoherent equilibration time, and hence does not contribute to the physical picture. We can thus conclude that in this region the many-body model only shows evolution of neutrino flavour equal to that found by the one-body approximation.

In the region $|N - M| > \sqrt{2(M + N)}$, the system exhibits fast oscillations due to semi-classical behaviour in which the quantum-mechanically fluctuating x and y components of the net angular momentum are negligible. Equilibration then occurs due to the quantum x and y components becoming large. The fast oscillations are not important because the amplitude is very small. For this same reason the periodicity can also be removed from the analysis. The small amplitude is interpreted as a freeze-out of the system in which little evolution takes place. We can conclude that in this regime the flavour evolution is slower than that predicted by the one-body description which takes into account all incoherent scattering. Thus the one-body prediction is also valid in this regime.

# Chapter 5

# Probability Correlation

The properties of the probability of one of the initial spin up particles remaining in the spin up state have been surprising and the results have shown a diverse range of interesting physical consequences . It was found that the behaviour of the probability changed drastically according to the initial configuration of the spins i.e. the number of spin up particles and the number of spin down particles. To further explore the system we have constructed and analysed the two body correlation function for a few initial conditions. This two body correlation function is interpreted as a correlation between probabilities rather than spatial correlations because the omission of the momentum variable is equivalent to ignoring the spatial location of the neutrinos.

## 5.1   The General Correlation Function

The two body correlation function is defined as,

$$\Gamma = \langle s_1 s_2 \rangle - \langle s_1 \rangle \langle s_2 \rangle \tag{5.1}$$

Here $\langle s_i \rangle$ is the expectation value of the angular momentum $s_i$, where $i = 1, 2$ refers to particle i.

Some definitions will be needed for the next section to increase readability. They are,

$P(\uparrow_i)$ is the probability that particle i is in the state up

$P(\downarrow_i)$ is the probability that particle i is in the state down

$P(\uparrow_i \downarrow_j)$ is the probability that particle i ($\neq j$) is in the state up and particle j is in the state down

$P(\downarrow_i \uparrow_j)$ is the probability that particle i ($\neq j$) is in the state down and particle j is in the state up

$P(\uparrow_i \uparrow_j)$ is the probability that both particle i and j are in the state up

$P(\downarrow_i \downarrow_j)$ is the probability that both particle i and j are in the state down.

The various parts of the correlation function are shown below.

$$\langle s_1 \rangle = \frac{1}{2}\left[P(\uparrow_1) - P(\downarrow_1)\right]. \tag{5.2}$$

$$= P(\uparrow_1) - \frac{1}{2}. \tag{5.3}$$

Note that $\langle s_2 \rangle$ is the same as Eqs. (5.2,5.3) but with the subscript changed to 2.

$$\langle s_1 s_2 \rangle = \frac{1}{4}\left[P(\uparrow_1 \uparrow_2) + P(\downarrow_1 \downarrow_2) - P(\uparrow_1 \downarrow_2) - P(\downarrow_1 \uparrow_2)\right]. \tag{5.4}$$

$$= \frac{1}{2}\left[P(\uparrow_1 \uparrow_2) + P(\downarrow_1 \downarrow_2)\right] - \frac{1}{4} \tag{5.5}$$

where Eq.(5.4) follows from the unit sum of the probabilities for all of the two spin configurations.

Using Eqs (5.2, 5.4) the general two body correlation function is,

$$
\begin{aligned}
\Gamma \;=\; & \frac{1}{4}\left\{[P(\uparrow_1\uparrow_2) - P(\uparrow_1)P(\uparrow_2)] + [P(\downarrow_1\downarrow_2) - P(\downarrow_1)P(\downarrow_2)]\right\} \\
& -\; \frac{1}{4}\left\{[P(\uparrow_1\downarrow_2) - P(\uparrow_1)P(\downarrow_1)] + [P(\downarrow_1\uparrow_2) - P(\downarrow_1)P(\uparrow_2)]\right\}.
\end{aligned}
\tag{5.6}
$$

The correlation function will be calculated for two different scenarios. The first is with an initial condition for which particle 1 and particle 2 are in the state up, which we will call initial condition 1. The second is with an initial condition for which particle 1 is in the state up and particle 2 is in the state down, which we call initial condition 2.

## 5.2   Correlation function with initial condition 1

We construct a correlation function with the first particle being in the state up initially and the second particle being in the same state initially. We use Eqs (5.3, 5.5) to construct the correlation function for ease of computation (as there is only a few probabilities to construct in this expression),

$$
\Gamma = \frac{1}{2}[P(\uparrow_1\uparrow_2) + P(\downarrow_1\downarrow_2)] - P(\uparrow_1)P(\uparrow_2) + \frac{1}{2}[P(\uparrow_1) + P(\uparrow_2)] - \frac{1}{2}.
\tag{5.7}
$$

For this initial condition it is irrelevent which particle is which, hence we drop the subscripts 1 and 2, and find,

$$\Gamma = \frac{1}{2}[P(\uparrow\uparrow) + P(\downarrow\downarrow)] - P(\uparrow)^2 + P(\uparrow) - \frac{1}{2} \tag{5.8}$$

## 5.2.1 Construction of Probabilities

**The Probability of one of the particles remaining in the state up**

The probability, $P(\uparrow)$ has been constructed in chapter 3. Here we reproduce the result,

$$P_1(t) = \sum_{J=|j_N-j_M|}^{j_N+j_M} \sum_{j=J-\frac{1}{2}}^{J+\frac{1}{2}} C_\uparrow(J) + \sum_{J=|j_N-j_M|}^{j_N+j_M-1} \eta_\uparrow(J) \cos[t\Delta E(J+1,J)] \tag{5.9}$$

where,

$$\begin{aligned} C_\uparrow(J) = {} & (2j_N+1)(2j+1) \left| \langle\, j_N, j_M, J, m \mid j_N, j_M, m_N, m_M \,\rangle \right|^2 \\ & \times \left| \langle\, \frac{1}{2}, j, \frac{1}{2}, m - \frac{1}{2} \mid \frac{1}{2}, j; J, m \,\rangle \right|^2 \\ & \times \left| \left\{ \begin{matrix} \frac{1}{2} & j_N - \frac{1}{2} & j_N \\ j_M & J & j \end{matrix} \right\} \right|^2 \end{aligned} \tag{5.10}$$

and

$$\eta_\uparrow(J) = -2(2j_N + 1)(2J + 2)$$

$$\times \langle\, j_N, j_M, J, m \mid j_N, j_M, m_N, m_M \,\rangle$$

$$\times \langle\, j_N, j_M, J+1, m \mid j_N, j_M, m_N, m_M \,\rangle$$

$$\times \langle\, \frac{1}{2}, J + \frac{1}{2}, \frac{1}{2}, m - \frac{1}{2} \mid \frac{1}{2}, J + \frac{1}{2}; J, m \,\rangle \qquad (5.11)$$

$$\times \langle\, \frac{1}{2}, J + \frac{1}{2}, \frac{1}{2}, m - \frac{1}{2} \mid \frac{1}{2}, J + \frac{1}{2}; J+1, m \,\rangle$$

$$\times \begin{Bmatrix} \frac{1}{2} & j_N - \frac{1}{2} & j_N \\ j_M & J & J + \frac{1}{2} \end{Bmatrix} \begin{Bmatrix} \frac{1}{2} & j_N - \frac{1}{2} & j_N \\ j_M & J + 1 & J + \frac{1}{2} \end{Bmatrix}.$$

**Probability of particle 1 and particle 2 remaining in the state up**

The probability of particle 1 and particle 2, which are initially in the state up, remaining in the state up is found by constructing a wavefunction and then a density matrix.

We begin with the wavefunction in the total angular momentum basis, J. The definitions of the variables are the same as previously. We also use the abbreviated form of the Clebsch-Gordan coefficients unless the full form is required for clarity.

$$\langle\, j_1 j_2 m_1 m_2 \mid j_1 j_2 Jm \,\rangle \equiv \langle\, j_1 j_2 m_1 m_2 \mid Jm \,\rangle. \qquad (5.12)$$

The wavefunction in the total angular momentum basis is,

$$\Psi(t) = \mid j_N m_N \,\rangle \otimes \mid j_M m_M \,\rangle = \sum_J e^{igtJ(J+1)} \langle\, Jm \mid j_N j_M m_N m_M \,\rangle \mid Jm \,\rangle. \qquad (5.13)$$

It is necessary to recouple the angular momenta to construct the probability. To do this we use a 6-j coefficient to make it explicit that particle 1, with angular momentum $s_1 = \frac{1}{2}$, and particle 2, with angular momentum $s_2 = \frac{1}{2}$, contribute to forming the angular momentum, $j_N$. This ensures that particle 1 and particle 2 are in the state up initially. Further the 6-j coefficient groups $s_1$ and $s_2$ into the angular momentum $s = 1$ and the left over angular momentum of all the spin up particles $j_N - 1$ and $j_M$ into the angular momentum $j_3$. The recoupling changes the wavefunction to,

$$
\begin{aligned}
\Psi(t) = \quad & \sum_J \quad (-1)^{-(j_N + j_M + J)} \sqrt{(2j_N + 1)(2j_3 + 1)} \langle\, Jm \mid j_N j_M m_N m_M \,\rangle \\
& \times \quad \begin{Bmatrix} 1 & j_N - 1 & j_N \\ j_M & J & j_3 \end{Bmatrix} e^{igtJ(J+1)} \mid 1 j_3 Jm \,\rangle .
\end{aligned}
\tag{5.14}
$$

Changing basis,

$$
\begin{aligned}
\Psi(t) = \quad & \sum_J \sum_{m_s} \sum j_3 \quad (-1)^{-(j_N + j_M + J)} \sqrt{(2j_N + 1)(2j_3 + 1)}\; e^{igtJ(J+1)} \\
& \times \qquad \langle\, Jm \mid j_N j_M m_N m_M \,\rangle \langle\, Jm \mid 1\, j_3\, m_s\, m - m_s \,\rangle \\
& \times \qquad \begin{Bmatrix} 1 & j_N - 1 & j_N \\ j_M & J & j_3 \end{Bmatrix} \mid 1\, m_s \,\rangle \otimes \mid j_3\, m - m_s \,\rangle .
\end{aligned}
\tag{5.15}
$$

The summation limits are as follows,

$$
\begin{aligned}
& |j_N - j_M| < J < j_N + j_M \\
& -1 < m_s < 1 \\
& |j_N - j_M - 1| < j_3 < j_N + j_M - 1 .
\end{aligned}
\tag{5.16}
$$

Note also that according to the Clebsch-Gordan $\langle\, Jm \mid 1\, j_3\, m_s\, m - m_s \,\rangle$ an alternate summation limit for $j_3$ is,

$$|J - 1| < j_3 < J + 1. \tag{5.17}$$

It is now possible to construct a two body density matrix by using the definition,

$$\rho_2(t) = |\Psi(t)\rangle\langle\Psi(t)|, \tag{5.18}$$

Eq. (5.15), and tracing over the variable $j_3$,

$$
\begin{aligned}
\rho_2 = \ & \sum_{J,J'} \sum_{m_s,m_s'} \sum_{j_3} \quad (-1)^{J-J'}(2j_N+1)(2j_3+1)\, e^{igt[J(J+1)-J'(J'+1)]} \\
& \times \quad \langle Jm \mid j_N j_M m_N m_M \rangle\langle J'm \mid j_N j_M m_N m_M \rangle \\
& \times \quad \langle Jm \mid 1\, j_3\, m_s\, m - m_s \rangle\langle J'm \mid 1\, j_3\, m_s'\, m - m_s' \rangle \\
& \times \quad \left\{\begin{matrix} 1 & j_N-1 & j_N \\ j_M & J & j_3 \end{matrix}\right\} \left\{\begin{matrix} 1 & j_N-1 & j_N \\ j_M & J' & j_3 \end{matrix}\right\} \\
& \times \quad |1\, m_s\rangle\langle 1\, m_s'|. \tag{5.19}
\end{aligned}
$$

To find the probability that particle 1 and 2 are in the state up we change basis one more time,

$$
\begin{aligned}
\rho_2 = \quad & \sum_{J,J'} \sum_{m_s,m'_s} \sum_{m_{s_1},m'_{s_1}} \sum_{m_{s_2},m'_{s_2}} \sum j_3 \quad (-1)^{J-J'}(2j_N+1)(2j_3+1) \, e^{igt[J(J+1)-J'(J'+1)]} \\
\times \quad & \langle \, Jm \mid j_N j_M m_N m_M \, \rangle \langle \, J'm \mid j_N j_M m_N m_M \, \rangle \\
\times \quad & \langle \, Jm \mid 1 \, j_3 \, m_s \, m-m_s \, \rangle \langle \, J'm \mid 1 \, j_3 \, m'_s \, m-m'_s \, \rangle \\
\times \quad & \langle \, s_1 s_2 \, m_{s_1} m_{s_2} \mid 1 \, m_s \, \rangle \langle \, s'_1 s'_2 \, m'_{s_1} m'_{s_2} \mid 1 \, m'_s \, \rangle \\
\times \quad & \left\{ \begin{matrix} 1 & j_N-1 & j_N \\ j_M & J & j_3 \end{matrix} \right\} \left\{ \begin{matrix} 1 & j_N-1 & j_N \\ j_M & J' & j_3 \end{matrix} \right\} \\
\times \quad & \mid s_1 s_2 \, m_{s_1} m_{s_2} \, \rangle \langle \, s'_1 s'_2 \, m'_{s_1} m'_{s_2} \mid. \qquad (5.20)
\end{aligned}
$$

Now setting $s_1 = s'_1 = \frac{1}{2}$, $s_2 = s'_2 = \frac{1}{2}$, $m_{s_1} = m'_{s_1} = \frac{1}{2}$ and $m_{s_2} = m'_{s_2} = \frac{1}{2}$ gives us the probability of particle 1 and particle 2 having projection $+\frac{1}{2}$ as the operator for the z component of the spin is diagonal. Note that equating the dashed and undashed variables ensures that we are only dealing with the diagonal components of the density matrix which are probabilities. After some algebra,

$$
\begin{aligned}
P(\uparrow\uparrow) = \ & \sum_{J=|j_N-j_M|}^{j_N+j_M} \sum_{j_3=J-1}^{J+1} \ (2j_N+1)(2j_3+1)|\langle\, Jm \mid 1\, j_3\, 1\, m-1\,\rangle|^2 \\
& |\langle\, Jm \mid j_N j_M m_N m_M\,\rangle|^2 \left|\left\{ \begin{array}{ccc} 1 & j_N-1 & j_N \\[2mm] j_M & J & j_3 \end{array} \right\}\right|^2 \\[4mm]
& + \sum_{J=|j_N-j_M|}^{j_N+j_M-1} \sum_{j3=J}^{J+1} \ 2(2j_N+1)(2j_3+1)\cos[gt(2J+2)] \\
& \langle\, Jm \mid j_N j_M m_N m_M\,\rangle\langle\, J+1\, m \mid j_N j_M m_N m_M\,\rangle \\
& \langle\, Jm \mid 1\, j_3\, 1\, m-1\,\rangle\langle\, J+1\, m \mid 1\, j_3\, 1\, m-1\,\rangle \\
& \left\{ \begin{array}{ccc} 1 & j_N-1 & j_N \\[2mm] j_M & J & j_3 \end{array} \right\}\left\{ \begin{array}{ccc} 1 & j_N-1 & j_N \\[2mm] j_M & J+1 & j_3 \end{array} \right\} \\[4mm]
& + \sum_{J=|j_N-j_M|}^{j_N+j_M-2} \ 2(2j_N+1)(2J+3)\cos[gt(4J+6)] \\
& \langle\, Jm \mid j_N j_M m_N m_M\,\rangle\langle\, J+2\, m \mid j_N j_M m_N m_M\,\rangle \\
& \langle\, Jm \mid 1\, J+1\, 1\, m-1\,\rangle\langle\, J+2\, m \mid 1\, J+1\, 1\, m-1\,\rangle \\
& \left\{ \begin{array}{ccc} 1 & j_N-1 & j_N \\[2mm] j_M & J & J+1 \end{array} \right\}\left\{ \begin{array}{ccc} 1 & j_N-1 & j_N \\[2mm] j_M & J+2 & J+1 \end{array} \right\}
\end{aligned}
$$

$$\tag{5.21}$$

## 5.2.2 Probability of particle 1 and particle 2 being in the state down

The probability of particle one and particle two, which are initially in the state up, being in the state down at some later time is constructed analogously to the previous section. The expression for the probability is the same as Eq. (5.21) but with the following changes,

$$m - 1 \rightarrow m + 1$$

$$1 \rightarrow -1 \text{ (this switch applies only to projections of angular momenta).}$$

$$(5.22)$$

### 5.2.3 Analysis

The correlation function was found by substituting Eq. (5.21), Eq. (5.9) and the probability for particle 1 and 2 being in the state down into Eq. (5.23). To analyse the correlation function in time we have plotted it for various numbers of spin up particles, N, and spin down particles, M. A subset of these simulations is shown in figure 5.1 and figure 5.2 for which $N = 101$ and M is varied.

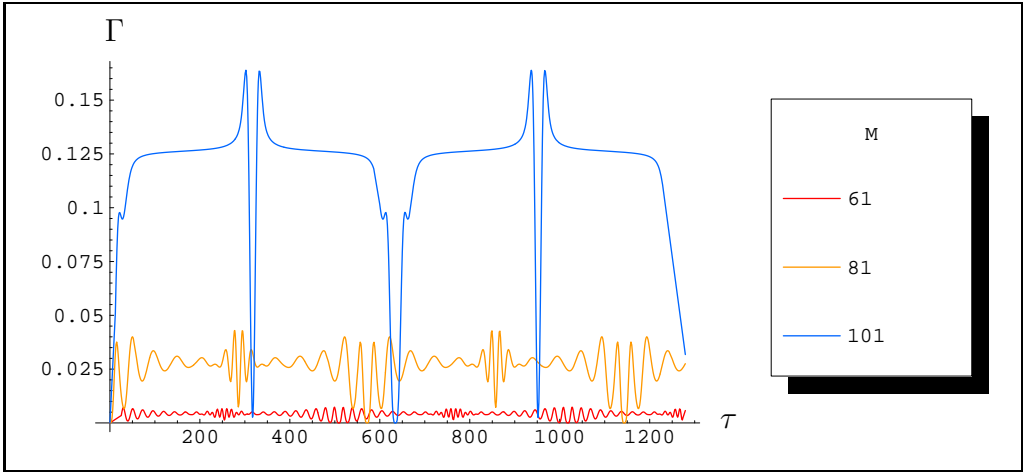

Figure 5.1: A plot of the correlation function, $\Gamma$, using initial conditon 1 in which particle 1 and particle 2 are in the state up. In these plots $N = 101$, M is varied such that $N \geq M$ . The time is scaled as usual, $\tau = gt(N + M)$. Note that the correlation function is positive for all cases. For the case $N = M$ the glitches in the plot are not real, rather they are an artifact of mathematica graphics

The correlation function shows the same features as those found in chapter 4 for the probability of one of the initial spin up particles remaining in the spin up state.

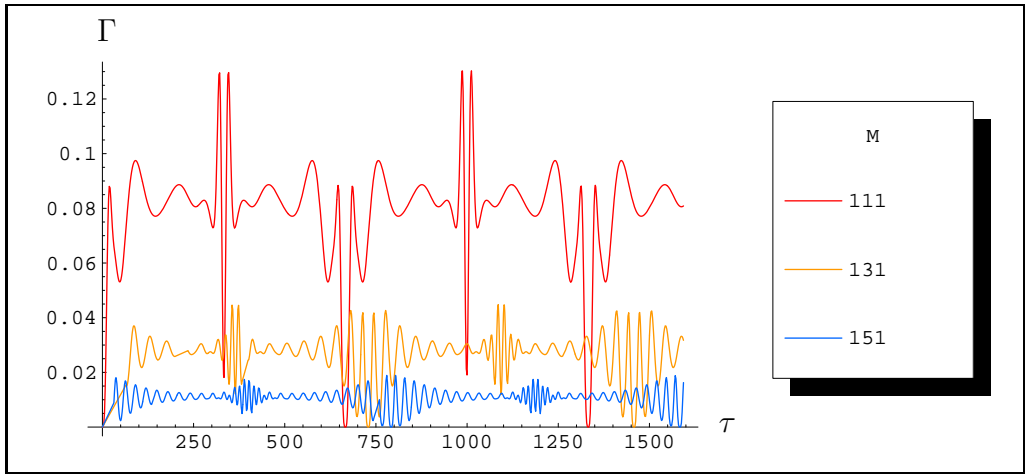

Figure 5.2: A plot of the correlation function, $\Gamma$, using initial condition 1 in which particle 1 and particle 2 are in the state up. In these plots $N = 101$ and M is varied such that $N < M$ . The time is scaled as usual, $\tau = gt(N + M)$. Note that the correlation function is positive for all cases.

Namely,

- When $N \sim M$, the system comes to equilibrium after some time. This features can be seen for the case N=101, M=101.

- When $N - M$ is large the correlation function exhibits oscillations but the amplitude is small.

- The correlation function is periodic.

However a new contribution to the analysis is that the correlation is positive. A discussion featuring this observation is deferred until section 5.5.

## 5.3  Correlation function with initial condition 2

In this section we calculate the correlation function for which the first particle is in state up and the second particle is in state down. We use Eqs (5.3, 5.5) to construct the correlation function for ease of computation (as there is only a few probabilities to construct in this expression),

$$\Gamma = \frac{1}{2}[P(\uparrow_1\uparrow_2) + P(\downarrow_1\downarrow_2)] - P(\uparrow_1)P(\uparrow_2) + \frac{1}{2}[P(\uparrow_1) + P(\uparrow_2)] - \frac{1}{2}. \qquad (5.23)$$

Unlike the correlation function with initial condition 1 the labels here are important as each particle begins in a different state.

### 5.3.1  Calculation of Probabilites

**The Probability of a particle being in the state up**

The probability of particle 1 (which is in the state up initally) remaining in the state up is as in Eq. (5.9) The probability of particle 2 (which is in the state down initially) being in the state up at some later time can be found in a similar fashion to that of the probability of particle 1. To begin the derivation of the probability of particle 2 being in the state down we write down the wavefunction in the total angular momentum basis, J,

$$\Psi(t) = \mid j_N m_N \;\rangle \otimes \mid j_M m_M \;\rangle = \sum_J e^{igtJ(J+1)} \langle\; Jm \mid j_N j_M m_N m_M \;\rangle \mid Jm \;\rangle. \qquad (5.24)$$

Since we are interested in finding particle 2 in the state down initally, we need to recouple. The total angular momentum is now formed by adding the angular mo-mentum of one of the spin down particles and the rest of the $N + M - 1$ particles. The angular momentum of the $N + M - 1$ particles, q, is the addition of the angular momentum of all the spin up particles, $j_N$ and the angular momentum of $M - 1$ spin down particles, $l = j_M - \frac{1}{2}$. Figure 5.3 represents this recoupling.

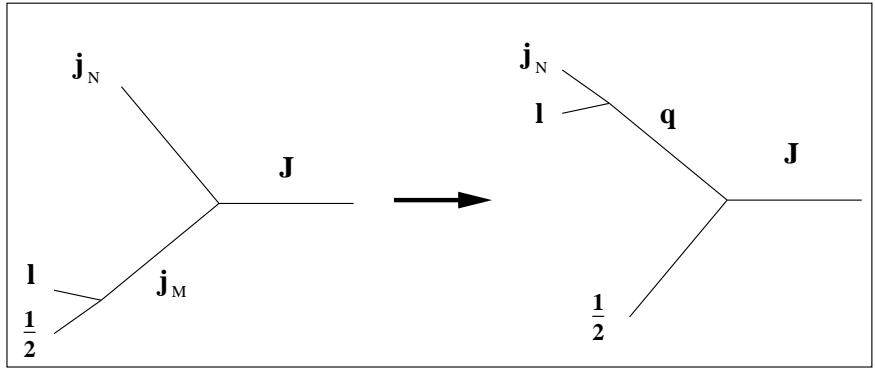

Figure 5.3: The left side of the figure represents the original coupling where we specify that the angular momenta $j_M$ was formed by adding the angular momentum $\frac{1}{2}$ with projection $-\frac{1}{2}$ and the angular momenta of $M - 1$ spin down particles, $l = j_M - \frac{1}{2}$. The right side of the figure represents the new coupling where we have seperated one of the spin down particles with angular momentum $\frac{1}{2}$.

Hence we arrive at,

$$
\begin{aligned}
\Psi_2(t) \quad &= \sum_J \sum_j \quad (-1)^{j_N + j_M + J} [(2q+1)(2j_M+1)]^{\frac{1}{2}} e^{igtJ(J+1)} \\
&\quad \langle\, Jm \mid j_N j_M m_N m_M \,\rangle \\
&\quad \begin{Bmatrix} j_N & j_M - \frac{1}{2} & q \\ \frac{1}{2} & J & j_M \end{Bmatrix} \mid q\, \frac{1}{2}\, J\, m \,\rangle.
\end{aligned}
\tag{5.25}
$$

The 6-j coefficient in the above equation represents the recoupling.

Further, changing basis to a product space of the angular momentum of the spin half particle and all the other particles,

$$
\begin{aligned}
\Psi_2(t) \quad =& \sum_J \sum_j \quad (-1)^{j_N+j_M+J}[(2q+1)(2j_M+1)]^{\frac{1}{2}} e^{igtJ(J+1)} \\
& \langle\, Jm \mid j_N j_M m_N m_M\,\rangle \langle\, Jm \mid q\,\frac{1}{2}\,m-\alpha\,\alpha\,\rangle \\
& \left\{ \begin{array}{ccc} j_N & j_M-\frac{1}{2} & q \\ \frac{1}{2} & J & j_M \end{array} \right\} \mid q\,m-\alpha\,\rangle \otimes \mid \frac{1}{2}\,\rangle.
\end{aligned} \tag{5.26}
$$

Here $\alpha$ is the projection of the spin half particle. We construct a density matrix from this wavefunction and use the same techniques as for the probaiblity of an inital spin up partcile preserving this state at a later time to find,

$$
P(\uparrow_2) = \sum_{J=|j_N-j_M|}^{j_N+j_M} \sum_{q=J-\frac{1}{2}}^{J+\frac{1}{2}} C_\downarrow(J) + \sum_{J=|j_N-j_M|}^{j_N+j_M-1} \eta_\downarrow(J) \cos[t\Delta E(J+1,J)] \tag{5.27}
$$

where,

$$
\begin{aligned}
C_\downarrow(J) =& (2j_N+1)(2j+1)\, |\langle\, j_N, j_M, J, m \mid j_N, j_M, m_N, m_M\,\rangle|^2 \\
& \times \left| \langle\, q, \frac{1}{2}, m-\frac{1}{2}, \frac{1}{2} \mid q, \frac{1}{2}; J, m\,\rangle \right|^2 \\
& \times \left| \left\{ \begin{array}{ccc} j_N & j_M-\frac{1}{2} & q \\ \frac{1}{2} & J & j_M \end{array} \right\} \right|^2
\end{aligned} \tag{5.28}
$$

and

$$\eta_{\downarrow}(J) = -2(2j_M + 1)(2J + 2)$$

$$\times \langle\, j_N, j_M, J, m \mid j_N, j_M, m_N, m_M \,\rangle$$

$$\times \langle\, j_N, j_M, J+1, m \mid j_N, j_M, m_N, m_M \,\rangle$$

$$\times \langle\, J+\frac{1}{2}, \frac{1}{2}, m-\frac{1}{2}, \frac{1}{2} \mid J+\frac{1}{2}, \frac{1}{2}; J, m \,\rangle \tag{5.29}$$

$$\times \langle\, J+\frac{1}{2}, \frac{1}{2}, m-\frac{1}{2}, \frac{1}{2} \mid J+\frac{1}{2}, \frac{1}{2}; J+1, m \,\rangle$$

$$\times \left\{ \begin{matrix} j_N & j_M - \frac{1}{2} & J + \frac{1}{2} \\ \frac{1}{2} & J & j_M \end{matrix} \right\} \left\{ \begin{matrix} j_N & j_M - \frac{1}{2} & J + \frac{1}{2} \\ j_M & J + 1 & j_M \end{matrix} \right\}.$$

**The probability of particle 1 (initially in the up state) and particle 2 (initially in the down state) being in the up state**

The method for constructing the probability of an initial spin up particle (1) and the initial spin down particle (2) both being in the spin up state at a later time is similar to that of section 5.2.1. However the recoupling here leads to a 9-j coefficient instead of a 6-j coefficient.

The recoupling scheme required here is illustrated in figure 5.4

The above diagram translates to the following 9-j coeffient,

$$\langle\, (s_1 l_1) j_N \,(s_2 l_2) j_M \, J \mid (s_1 s_2) s \,(l_1 l_2) L \,\rangle \equiv [(2j_N + 1)(2j_M + 1)(2s + 1)(2L + 1)]^{\frac{1}{2}} \left\{ \begin{matrix} s_1 & l_1 & j_N \\ s_2 & l_2 & j_M \\ s & L & J \end{matrix} \right\} \tag{5.30}$$

Note that $s_1$ is one of the spin particles that makes up the angular momenta $j_N$. Hence particle 1 with angular momentum $s_1$ is in the state up initially. Similarly, $s_2$ is one of the particles that make up the angular momenta $j_M$. Hence particle 2 with angular momentum $s_2$ is in the state down initially.

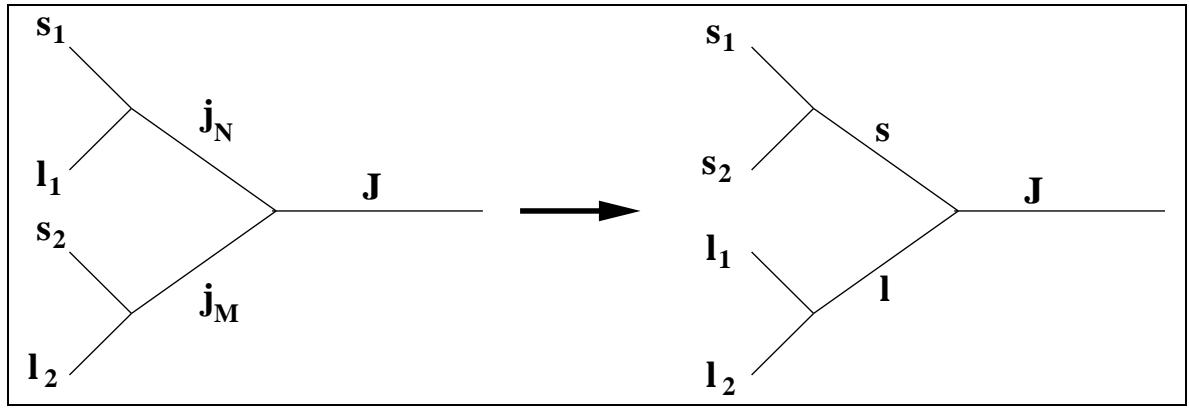

Figure 5.4: The left side of this figure represents the original coupling and the right side represents the new coupling

The angular momenta in the 9-j coefficient are,

$$s_1 \quad = \quad \frac{1}{2} \tag{5.31}$$

$$s_2 \quad = \quad \frac{1}{2} \tag{5.32}$$

$$l_1 \quad = \quad j_N - \frac{1}{2} \tag{5.33}$$

$$l_2 \quad = \quad j_M - \frac{1}{2} \tag{5.34}$$

$$|s_1 - s_2| \quad < s < \quad s_1 + s_2 \tag{5.35}$$

$$|l_1 - l_2| \quad < L < \quad l_1 + l_2. \tag{5.36}$$

By using a density matrix to find the probability of both particles being the up state at time t, we find,

$$
\begin{aligned}
P(\uparrow_1\uparrow_2) = \quad & \sum_{J=|j_N-j_M|}^{j_N+j_M} \sum_{L=J-1}^{J+1} 3(2j_N+1)(2j_M+1)(2L+1)|\langle\, Jm\mid 1\,L\,1\,m-1\,\rangle|^2 \\
& |\langle\, Jm\mid j_Nj_Mm_Nm_M\,\rangle|^2 \left|\left\{\begin{array}{ccc} \frac{1}{2} & j_N-\frac{1}{2} & j_N \\ \frac{1}{2} & j_M-\frac{1}{2} & j_M \\ 1 & L & J \end{array}\right\}\right|^2 \\[1em]
+ & \sum_{J=|j_N-j_M|}^{j_N+j_M-1} \sum_{L=J}^{J+1} 6(2j_N+1)(2j_M+1)(2jL+1)\cos[gt(2J+2)] \\
& \langle\, Jm\mid j_Nj_Mm_Nm_M\,\rangle\langle\, J+1\,m\mid j_Nj_Mm_Nm_M\,\rangle \\
& \langle\, Jm\mid 1\,L\,1\,m-1\,\rangle\langle\, J+1\,m\mid 1\,L\,1\,m-1\,\rangle \\
& \left\{\begin{array}{ccc} \frac{1}{2} & j_N-\frac{1}{2} & j_N \\ \frac{1}{2} & j_M-\frac{1}{2} & j_M \\ 1 & L & J \end{array}\right\} \left\{\begin{array}{ccc} \frac{1}{2} & j_N-\frac{1}{2} & j_N \\ \frac{1}{2} & j_M-\frac{1}{2} & j_M \\ 1 & L & J+1 \end{array}\right\} \\[1em]
+ & \sum_{J=|j_N-j_M|}^{j_N+j_M-2} 6(2j_N+1)(2j_M+1)(2J+3)\cos[gt(4J+6)] \\
& \langle\, Jm\mid j_Nj_Mm_Nm_M\,\rangle\langle\, J+2\,m\mid j_Nj_Mm_Nm_M\,\rangle \\
& \langle\, Jm\mid 1\,J+1\,1\,m-1\,\rangle\langle\, J+2\,m\mid 1\,J+1\,1\,m-1\,\rangle \\
& \left\{\begin{array}{ccc} \frac{1}{2} & j_N-\frac{1}{2} & j_N \\ \frac{1}{2} & j_M-\frac{1}{2} & j_M \\ 1 & J+1 & J \end{array}\right\} \left\{\begin{array}{ccc} \frac{1}{2} & j_N-\frac{1}{2} & j_N \\ \frac{1}{2} & j_M-\frac{1}{2} & j_M \\ 1 & J+1 & J+2 \end{array}\right\}
\end{aligned}
\tag{5.37}
$$

**The probability of particle 1(initially in the up state) and particle 2 (initially in the down state) being in the state down**

The probability that particle 1, initially in the up state, and particle 2, initially in the down state, are in the upstate at a later time is the same as Eq. (5.37) but with the following changes,

$$m - 1 \rightarrow m + 1$$

$$1 \rightarrow -1 \text{ (this switch applies only to projections of angular momenta)}. \tag{5.38}$$

### 5.3.2 Analysis

The correlation function was found by substituting Eqs. (5.37), (5.9), (5.27) and the equation for the probability that particle 1 (initially up) and particle 2 (initially down) are in the up state into the correlation expression, Eq.(5.23). Figure 5.5 and figure 5.6 show plots of the correlation function for $N = 101$ and various numbers of M.

The features of this correlation function are similar to those found in the correlation function with initial condition 1 and that of chapter 4. That is, the correlation function displays equilibration, freeze-out and periodicity. The difference is that the correlation function is negative. The discussion of this feature is deferred until section 5.5.

## 5.4 A short note on the period

The correlation function is periodic. Indeed this is expected as the correlation function is just a finite sum of trigonometric functions. This period of the correlation function for the two cases that we have studied is the same as that found in chapter 4, Eq.(4.55). The calculation of the period is similar to that shown in chapter 4 and can be found

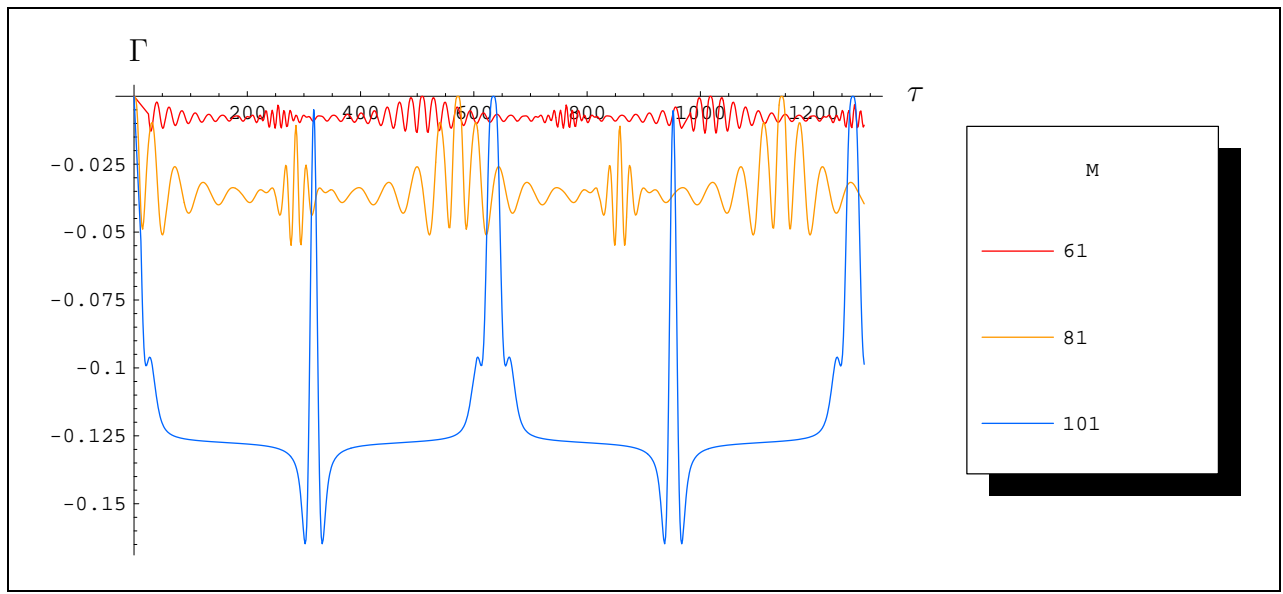

Figure 5.5: A plot of the correlation function, $\Gamma$, using initial conditon 2 in which particle one is in the state up and particle 2 is in the state down. In these plots $N = 101$, M is varied such that $N \geq M$ . The time is scaled as usual, $\tau = gt(N + M)$. Note that the correlation function is negative for all cases. For the case $N = M$ the glitches in the plot are not real, rather they are an artifact of mathematica graphics

in Appendix A.

## 5.5  Discussion

By analysing the correlation function for two cases we have found two different behaviours: for initial condition 1, the correlation function is positive and for initial condition 2 the correlation function is negative. Let us remind the reader of the expression for the general correlation function ,

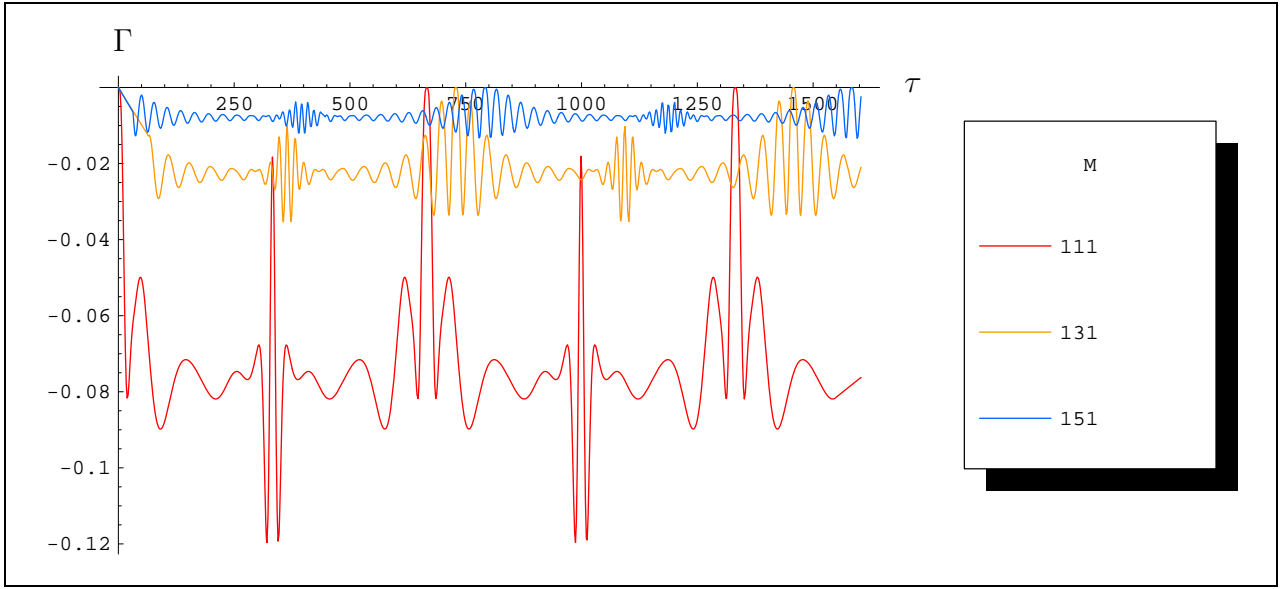

Figure 5.6: A plot of the correlation function, $\Gamma$, using initial condiiton 2 in which particle one is in the state up and particle 2 is in the state down. In these plots $N = 101$ and M is varied such that $N < M$ . The time is scaled as usual, $\tau = gt(N + M)$. Note that the correlation function is negative for all cases.

$$
\begin{aligned}
\Gamma \;=\; & \frac{1}{4} \left\{ [P(\uparrow_1 \uparrow_2) - P(\uparrow_1) P(\uparrow_2)] + [P(\downarrow_1 \downarrow_2) - P(\downarrow_1) P(\downarrow_2)] \right\} \\
- \;& \frac{1}{4} \left\{ [P(\uparrow_1 \downarrow_2) - P(\uparrow_1) P(\downarrow_1)] - [P(\downarrow_1 \uparrow_2) - P(\downarrow_1) P(\uparrow_2)] \right\} .
\end{aligned}
\tag{5.39}
$$

Let us define,

$$
\delta(s_1, s_2) \equiv [P(s_1, s_2) - P(s_1) P(s_2)] ,
\tag{5.40}
$$

where $s_1$ is the initial state of particle 1 and $s_2$ is the initial state of particle two.

The correlation function in Eq.(5.39), rewritten with the new definitions, is,

$$
\Gamma = \frac{1}{4} \left[ \delta(\uparrow_1, \uparrow_2) + \delta(\downarrow_1, \downarrow_2) - \delta(\uparrow_1, \downarrow_2) - \delta(\downarrow_1, \uparrow_2) \right] .
\tag{5.41}
$$

Recall that the first case involved specifying that the inital conditions for particle 1 and particle 2 were identical. That is both particles were in the state up initially. This led to a correlation which was always positive. The second initial condition specified that particle 1 and particle two were aligned opposite to each other. That is, particle 1 was in the state up initially and particle 2 was in the state down initially. Here we found the correlation was always negative. What is the physical meaning of this?

Note that

$$\sum_{s_1,s_2} \delta(s_1, s_2) = 0. \tag{5.42}$$

Therefore Eq.(5.41) becomes,

$$\Gamma = \frac{1}{2} \left[ \delta(\uparrow_1, \uparrow_2) + \delta(\downarrow_1, \downarrow_2) \right]. \tag{5.43}$$

Although we were unable to show the following analytically, numerically, for all case considered in figures 5.1, 5.2, 5.5 and 5.6, we have found,

$$\delta(\uparrow_1, \uparrow_2) - \delta(\downarrow_1, \downarrow_2) = 0. \tag{5.44}$$

With this property,

$$\Gamma = \delta(\uparrow_1, \uparrow_2). \tag{5.45}$$

Hence when the correlation function is positive,

$$\delta(\uparrow_1, \uparrow_2) > 0, \tag{5.46}$$

so,

$$P(\uparrow_1, \uparrow_2) > P(\uparrow_1)P(\uparrow_2) \tag{5.47}$$

Now consider the definition for conditional probability, $P(s_1|s_2)$, in which the state of particle 1 is $s_1$ given that particle 2 is in the state $s_2$,

$$P(s_1|s_2) = \frac{P(s_1, s_2)}{P(s_2)}.$$  (5.48)

Therefore we can rewrite Eq.(5.47) as,

$$P(\uparrow_1 \mid \uparrow_2) > P(\uparrow_1).$$  (5.49)

The interpretation of the positivity of the correlation function when particle 1 and particle 2 are both in the state up initially is now clear. The probability of particle 1 remaining in the state up is larger if particle 2 is in the state up also than the probability of particle 1 remaining in the spin up state if the state of particle 2 is unspecified. Therefore the spins prefer to be aligned in the same way as initially for all time.

When the correlation function is negative,

$$\delta(\uparrow_1, \uparrow_2) < 0,$$  (5.50)

and

$$P(\uparrow_1 \mid \uparrow_2) < P(\uparrow_1).$$  (5.51)

Therefore, in this case, the probability of particle one remaining in the state up if particle 2 is in the state up is less than the probability of particle one remaining in the state up if the state of particle 2 is not specified. Recall that the correlation function was negative when the initial state of particle 1 was up and the that of particle 2 was down. The result again points to the particles preferring to remain aligned in an opposite way because this was their initial configuration.

We can now conclude the system does not seem to exhibit an interesting time evolution. The construction of a two body density matrix required that we specify the initial conditions of two particles in the system rather than just one as in chapter 4. The knowledge of the initial state of the two particles has shown that the particles prefer to be aligned in the same way as they were initially, thus leading to very little evolution of the system. Hence one could conjecture that these results seems to show a pattern of the system displaying less and less dynamical time evolution the more we specify the initial conditions.

# Chapter 6

# The Simplified Many-Body Neutrino Model with Oscillations

In this chapter we are interested in adding neutrino oscillations to the simplified many-body neutrino model.

The phenomenon of neutrino oscillations can be included in the model by adding an extra term to the Hamiltonian that describes the oscillations. Recall that in our model neutrinos are equivalent to spins. Consider the following Hamiltonian for the ith spin,

$$H_i^{osc} = \vec{B} \cdot \vec{s}_i. \qquad (6.1)$$

The equation of motion for this spin is,

$$i\frac{d\vec{s}_i}{dt} = [H_i^{osc}, \vec{s}_i]$$
$$\frac{d\vec{s}_i}{dt} = \vec{B} \times \vec{s}_i. \qquad (6.2)$$

The vector B is thus an effective magnetic field which the spin, $\vec{s}_i$, precesses around. If we identify the effective magnetic field as $\vec{B} = \frac{\Delta m^2}{2E}(\sin 2\theta, 0, -\cos 2\theta)$, where $\theta$ is the mixing angle, then Eq.(6.2) represents neutrino oscillations. If we think of the z-axis as the flavour axis then the projection of $\vec{s}_i$ onto the z-axis gives the probability of finding the neutrino in a particular flavour state. As $\vec{s}_i$ precesses about $\vec{B}$ the projection of $\vec{s}_i$ onto the z-axis changes hence representing neutrino oscillations.

For the entire ensemble of neutrinos we must have,

$$\vec{J} = \sum_{i=1}^{M+N} \vec{s}_i, \qquad (6.3)$$

so that the neutrino oscillations Hamiltonian becomes,

$$H^{osc} = \vec{B} \cdot \vec{J}, \qquad (6.4)$$

with the equation of motion,

$$\frac{d\vec{J}}{dt} = \vec{B} \times \vec{J}. \qquad (6.5)$$

In this case, the total angular momentum, $\vec{J}$ precesses about the effective magnetic field, $\vec{B}$. Note that Eq.(6.5) is the well known spin-precession picture for vacuum neutrino oscillations [48, 49] .

The total Hamiltonian for our simplified many-body neutrino model with neutrino oscillation is thus,

$$H = \vec{B} \cdot \vec{J} + gJ^2 + g\frac{3}{4}(M+N)(M+N-2), \qquad (6.6)$$

with eigenvalues,

$$E(J, N, M) = BM_B + gJ(J+1) + g\frac{3}{4}(M+N)(M+N-2). \qquad (6.7)$$

Including neutrino oscillations in the Hamiltonian does not change our analysis drastically. Since the total angular momentum vector precesses about the effective magnetic field any results must first be projected on to this vector then the flavour axis (z-axis).

In the freeze-out regime the system showed no evolution. Recall that in the semi-classical picture the probability of one of the spin up particles remaining in this state can be written as,

$$P_1(t) = \frac{1}{2} + \frac{1}{N}\overline{j_{Nz}} \qquad (6.8)$$

The expectation value, $\overline{j_{Nz}}$ was found by projecting the angular momentum of all the spin up particles, $j_N$, onto the total angular momentum J (which precessed about the z-axis with $J_z$ being a constant) and then projecting onto the z-axis. It was found that in the freeze-out regime $\overline{j_{Nz}} \approx \frac{N}{2}$. With the inclusion of the effective magnetic field in the spin system or neutrino oscillations in the neutrino system, the z-component of the net angular momentum vector is no longer constant, so that the probability of one of the initial spin up particle remaining in the spin up state changes according to where J is in its trip around B. This is depicted in figure 6.1 where $\overline{j_{Nz}}$ oscillates between a and b. The probability therefore oscillates between $\frac{1}{2} + \frac{a}{N}$ and $\frac{1}{2} + \frac{b}{N}$ and the system is no longer frozen due to the effective magnetic field in the spin system or neutrino oscillations in the neutrino system.

Of course in a real neutrino system such as that of the early universe and supernovae the coherent scattering of neutrino off electrons also plays an important role. The MSW theory, discussed in chapter 2, predicts that a background of electrons could suppress the oscillations (as is the case near the supernova core) and thus equilibration and freeze-out could become the dominant effects. The effects of the background can also be included into our simplified model by adding the term $\sqrt{2}G_F\rho_e\hat{z}$ to the effective magnetic field, B. This has the effect of changing the angle between B and the z-axis. If the density of electrons is large then the angle between B and the z-axis is small, thus oscillations are suppressed and equilibration or freeze-out come out to play.

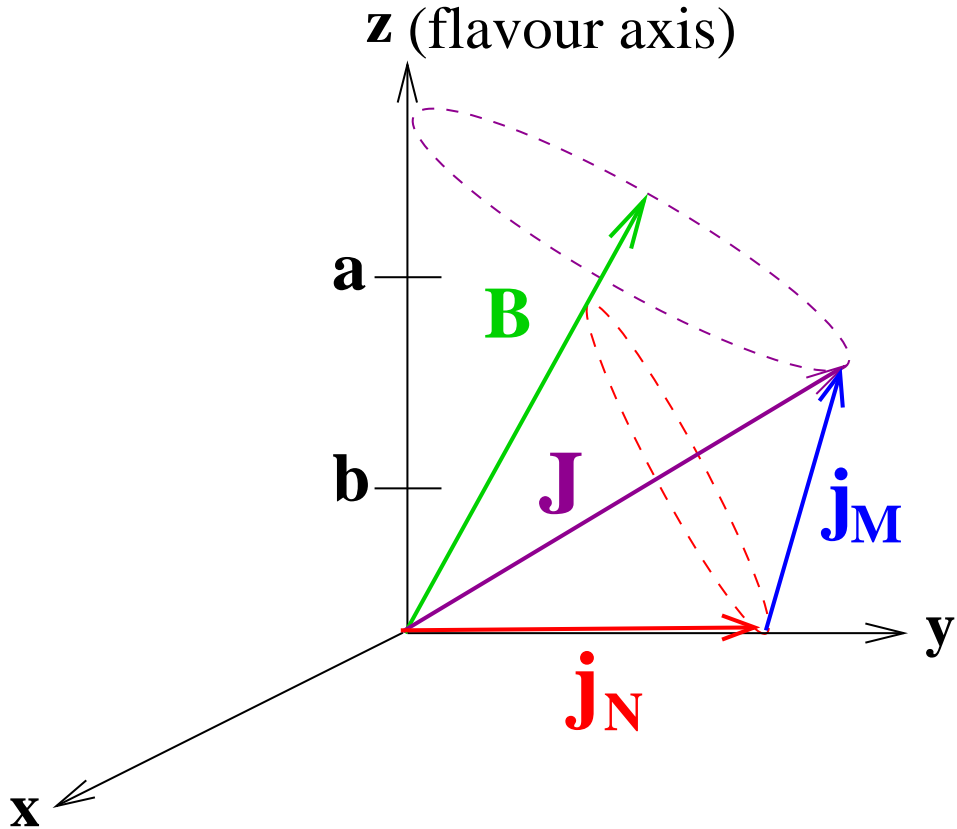

Figure 6.1: The inclusion of neutrino oscillations in the model introduces a new vector, B, which is an effective magnetic field that the total angular momentum, $J$, precesses around. The expectation value of $j_N$ now oscillates between the maximum value a and minimum value b. Therefore the probability also oscillates between the $\frac{1}{2} + \frac{a}{N}$ and $\frac{1}{2} + \frac{b}{N}$ in the freeze-out regime.

When necessary one can readily include physical effects of neutrino oscillations and matter background in detailed calculations analogous to those above.

# Chapter 7

# Conclusion

Throughout the history of physics a fundamental understanding of the universe has been gained by stripping the world to its bare essentials. The grand examples of such advances are Newton's theory with it's three laws of motion, Einstein's general relativity which describes matter and space in one equation and quantum mechanics governed by five postulates. Of course in these examples a great leap of imagination was also required but it was probably only possible after the details of the everyday world were removed. On the other hand, details need to be included in order to make use of the understanding in our world. Such examples can be found in the many feats of modern engineering: aeroplanes, televisions, cars, pacemakers etc. and in astrophysics, where, for example, it is essential to know the details to be able to explain how the big bang caused the universe we see today. In the modern world most of the technology we use must be described very accurately so that humans can benefit from the technology without harm. Theoretical physics models in the far, far away systems of the universe have no such prerequisite (well, not yet!), which is convenient because these complicated system are often too hard to be solved exactly. Many simplifica-tions and assumptions are needed to be able to solve and understand the equations that describe these systems. Two such systems involve neutrino-neutrino interactions: the early universe and supernova. Understandably any attempt at analysis has included the assumption that a one-body equation is sufficient in describing the dense neutrino ensemble. To address this assumption we have stripped the problem down to its bare essentials. Namely only looking at the forward scattering part of the Hamiltonian and removing the momentum degrees of freedom. With these simplifications and considering only equal strength interactions we were able to map the neutrino-neutrino interactions onto spin-spin interactions and solve the system exactly. The generalisation of the model constructed by F&L [4] in which there were equal numbers of spin up and spin down, proved to be instructive. Our analysis focused on the initial condition with N spin up particles and M spin down particles. In the neutrino system this is equivalent to neutrinos initially in flavour eigenstates. The solution revealed unexpected behaviour and interesting physics.

We have found that the system has two sets of different behaviours termed equilibration and freeze-out. In the case where $N \sim M$, the system came to equilibrium, in which no evolution took place, after a time proportional to the square root of the total number of particles. This time was identified with incoherent scattering in agreement with the predictions of the one-body description. In the other extreme, when $|N - M|$ was large, the system displayed fast coherent oscillations but with a very small amplitude. This was interpreted as the freeze-out of the system due to the semi-classical behaviour of the system. In the vector model of angular momentum the angular momentum with flactuating x and y components is represented as precessesing about the z-axis. In the case of the freeze-out the total angular momentum vector is pinned to z-axis (flavour axis) in which the quantum x and y components can be ignored. Once these x and y components become large equilibration is allowed to occur. This led to the conclusion that the fast coherent oscillations are due to semi-classical effects while the equilibration is due to quantumechanical effects. We have found that the boundary between equilibration and freeze-out is $|N - M| \sim \sqrt{2(N + M)}$. This shows that freeze-out is almost always present in the system unless $N \sim M$.

Further analysis of the system was performed by constructing a probability correlation function for two initial condition. This analysis was numerical and showed that the spins prefer to be aligned in the same way as initially for all time. From these results we conjectured that as we specify the initial condition of more and more particles the system seems to show less and less dynamical evolution.

Neutrino oscillations were also added to the system by including an effective magnetic field in the spin-spin interaction Hamiltonian. This resulted in the precession of the angular momentum vector about the magnetic field rather than the z-axis. Thus any results must be first projected onto the angular momentum vector then onto the z-axis. In particular, with the inclusion of the effective magnetic field, the system was no longer frozen when $|N - M|$ was large. The probability was able to oscillate according to the fluctuating value of the projection of the angular momentum vector onto the z-axis.

Finally, we note that the freeze-out effect is a new contribution to the field, in that it has not been found in any other analysis of the neutrino ensemble. Our model has included some simplifications which were outlined in chapter 3 and thus the freeze-out effect may only be an artifact of the simplifications employed when constructing the model. Hence further work could investigate whether the freeze-out effect is real in a neutrino gas. We must point out, however, that freeze-out does occur in the spin system. It is perhaps also worth pursuing adding the momentum degrees of freedom into the model so that it can describe a dense neutrino ensemble more realistically. In particular, this may be useful in studying neutrinos in the early universe.

The goal of this thesis was to ascertain whether the factorisation of the wavefunction into one-body states was valid in a dense neutrino system. We have shown that a many-body neutrino system, albeit with some simplifications, behaves in a way that is predicted by the one-body description. The extensive investigation has proved to contribute to a better understanding of a dense neutrino system. Further, the constructed system and analysis is valid for any two state system with equal strength interaction and thus we hope that our results will find applications beyond the neutrino field.

# Appendices

# Appendix A

# Calculation of the Period for the Correlation Function

Recall that the correlation function is,

$$\Gamma = \frac{1}{2}[P(\uparrow_1\uparrow_2) + P(\downarrow_1\downarrow_2)] - P(\uparrow_1)P(\uparrow_2) + \frac{1}{2}[P(\uparrow_1) + P(\uparrow_2)] - \frac{1}{2}. \tag{A.1}$$

and that the probabilities have a structure,

$$
\begin{align}
P(1) &= \sum_J C(J) + \sum_J \eta(J)\cos[2(J+1)gt] \tag{A.2}\\
P(1,2) &= \sum_J K(J) + \sum_J \alpha(J)\cos[2(J+1)gt] + \sum_J \beta(J)\cos[2(2J+3)gt] \tag{A.3}
\end{align}
$$

where 1 represents the initial state of particle 1 and 2 represents the initial state of particle 2. Note that $P(2)$ has the same structure as $P(1)$.

The method for determining the period is as follows. We find the period of each of the terms (or probabilities) in the correlation function. The period of the correlation function is then just the smallest period found.

## A.0.1   Period of P(1,2)

The period of P(1,2) is found by first finding the period of the second term of Eq.(A.3) analogously to that of section 4.4 in chapter 4 and then the period of the third term of Eq.(A.3) using the same method as that in chapter 4. It turns out that the the two periods are multiples of each other therefore the period of P(1,2) is the minimum of the two periods. We find,

$$T_{P(1,2)} = \begin{cases} \frac{\pi}{g} & \text{if N+M is even} \\ \frac{2\pi}{g} & \text{it N+M is odd} \end{cases} \tag{A.4}$$

## A.0.2   Period of P(1)P(2)

The probability $P(1)$ is just a cosine fourier series of the form,

$$P(t) = \sum_n c_n \cos[n\omega t] \tag{A.5}$$

where

$$\omega = \frac{2\pi}{T} \tag{A.6}$$

and T is the period of the cosine fourier series.

Now the term $P(\uparrow_1)P(\uparrow_2)$, is equivalent to,

$$P_1(t)P_2(t) = \sum_n \sum_m c_n d_m \cos[n\omega t \cos[m\omega t]]. \tag{A.7}$$

Now $\cos[x]\cos[y] = \frac{1}{2}[\cos[x+y] + \cos[x-y]]$, hence

$$P_1(t)P_2(t) = \sum_n \sum_m c_n d_m \frac{1}{2}\cos[(n+m)\omega t] + \sum_n \sum_m c_n d_m \frac{1}{2}\cos[(n-m)\omega t]. \quad (A.8)$$

Let $p = n + m$ in the first term of Eq.(A.8) and $s = n - m$ in the second term of Eq.(A.8),

$$P_1(t)P_2(t) = \frac{1}{2}\sum_p (\sum_n c_n d_{p-n}) \cos[p\omega t] + \frac{1}{2}\sum_s (\sum_n c_n d_{n-s}) \cos[s\omega t]. \quad (A.9)$$

By redefining the constants in the brackets as,

$$\kappa_p = \sum_n c_n d_{p-n} \quad (A.10)$$

$$\chi_s = \sum_n c_n d_{n-s} \quad (A.11)$$

and recognising that in a cosine Fourier series $d_{n-p} = d_{p-n}$, we find,

$$P_1(t)P_2(t) = \sum_p \kappa_p \cos[p\omega t]. \quad (A.12)$$

Hence the period of $P_1(t)P_2(t)$ is the same as $P_1(t)$. Therefore the period for $P(1)P(2)$ is,

$$T_{P(1)P(2)} = \begin{cases} \frac{\pi}{g} & \text{if } N+M \text{ is even} \\ \frac{2\pi}{g} & \text{if } N+M \text{ is odd} \end{cases} \quad (A.13)$$

### A.0.3   Period of the Correlation Function

We compare Eq.(A.4) and Eq.(A.13) to find that the period of the correlation function is,

$$T_\Gamma = \begin{cases} \frac{\pi}{g} & \text{if } N + M \text{ is even} \\[2ex] \frac{2\pi}{g} & \text{if } N + M \text{ is odd} \end{cases} \tag{A.14}$$